\documentclass[pre,superscriptaddress,showpacs]{revtex4}
\usepackage{amsmath,amsfonts,amssymb,graphicx,psfrag,bbm,bm}

\begin{document}
\title{Maxwell and very hard particle models for probabilistic ballistic annihilation: hydrodynamic description}
\author{Fran\c{c}ois Coppex}
\affiliation{Department of Theoretical Physics, University of Gen\`eve, CH-1211 Gen\`eve 4, Switzerland}
\author{Michel Droz}
\affiliation{Department of Theoretical Physics, University of Gen\`eve, CH-1211 Gen\`eve 4, Switzerland}
\author{Emmanuel Trizac}
\affiliation{Laboratoire de Physique Th\'eorique et Mod\`eles Statistiques (UMR 8626 du CNRS), B\^atiment 100, Universit\'e de Paris-Sud, 91405 Orsay, France}
\begin{abstract}
The hydrodynamic description of  probabilistic ballistic annihilation, for which no conservation laws hold, is an intricate problem with hard sphere-like dynamics for which no exact solution exists. We consequently focus on simplified approaches, the Maxwell and  very hard particles (VHP) models, which allows us to compute analytically upper and lower bounds for several quantities. The purpose is to test the possibility of describing such a far from equilibrium dynamics with simplified kinetic models. The motivation is also in turn to assess the relevance of some singular features appearing within the original model and the approximations invoked to study it. The scaling exponents are first obtained from the (simplified) Boltzmann equation, and are confronted against Monte Carlo simulation (DSMC technique). Then, the Chapman-Enskog method is used to obtain constitutive relations and transport coefficients. The corresponding Navier-Stokes equations for the hydrodynamic fields are derived for both Maxwell and VHP models.  We finally perform a linear stability analysis around the homogeneous solution, which illustrates the importance of dissipation in the possible development of spatial inhomogeneities.
\end{abstract}
\maketitle

%=========================================================
\section{Introduction}\label{section1}
%=========================================================

The possibility to describe in terms of hydrodynamic equations the evolution of a system where some physical quantities are not conserved  is a challenging problem of non-equilibrium statistical mechanics. Several questions have to be faced as for example the validity of the underlying (and in practice approximate) kinetic theory, the choice of the hydrodynamical fields that are supposed to describe the relevant excitations in the problem, or the consistency of the method itself that is used to deduce the coarse-grained description from the kinetic theory. Much attention has been recently paid to such questions, mainly in the field of granular gas dynamics (see e.g \cite{book,duftycondmat,new,duftydrey}). In such systems, the kinetic energy 
is not conserved, while the linear momentum and number of particles are. However, even for low dissipation, the derivation of the hydrodynamic relations, based on a hard-sphere-like Boltzmann equation is not a simple task and several approximations have to be invoked~\cite{duftycondmat}.  These difficulties lead to consider some simpler models by choosing ad-hoc collision term in the Boltzmann equation. The so-called Maxwell and very hard particles (VHP) models~\cite{vhp1,matthieu} are particularly interesting and reproduce some qualitative features of the granular gas of inelastic hard spheres~\cite{santos,Puglisi,aparna2,astillero,aparna1,garzomontanero,bennaim}.

Another class of problems for which not only energy but also the density and momentum are not conserved is probabilistic ballistic annihilation (PBA). In such a system, the particles move ballistically between collisions. When two particles meet, they undergo an instantaneous collision and are removed from the system with probability $p$ or undergo an elastic scattering with probability $ (1-p)$. Since collisions are assumed to be instantaneous, two body events only are taken into account. The PBA model was first introduced in one dimension in~\cite{richardson}. In the limit $p \to 0$, where density, momentum and kinetic energy are conserved, one recovers a system of hard spheres for which the hydrodynamic equations are well known~\cite{bogolubov,cohen2,chapmancowling}. The other limit $p=1$ (pure annihilation) has been the object of some work~\cite{elskens,BenNaim,piasecki,reydrozpiasecki,Blythe,krapivskysire,trizac,Lettre,physica2}. It was shown that in the long time limit the annihilation dynamics is exactly described by the Boltzmann equation in dimensions higher than one~\cite{trizac}. This may qualitatively be understood by the fact that the density of the gas decays and, at late times, the packing fraction is very low. This fact lead to conjecture that the Boltzmann equation is an adequate description of PBA at late times for $p>0$~\cite{coppex}.

Given that $p$ may be considered as a perturbation parameter allowing to recover the elastic limit, the PBA model is particularly interesting in view of testing the relevance and validity of the hydrodynamic description in general, which is a controversial issue and a long term goal of the present work . The analytical treatment with usual hard sphere dynamics however appears to be quite involved~\cite{coppexdroztrizac}, and we study here the simplified Maxwell and VHP versions of PBA. The motivation is here is not only to test the ability of simplified kinetic models to mimic the hard sphere dynamics for a model far from equilibrium (and with no conserved quantity, a more severe situation than that of granular gases) but also to shed some light on some peculiar features obtained in the hydrodynamic study of Ref.~\cite{coppexdroztrizac}. In particular, this work exhibited divergent transport coefficients for a critical value of $p$. We will see that such singularities are absent in the simplified approaches, which may indicate that they are not associated with any physically relevant phenomenon. It will also appear that Maxwell and VHP approaches provide useful bounds for the hard sphere dynamics, so that similar inequalities as those found in \cite{krapivskysire,trizacbound} concerning the scaling exponents can be obtained.

The paper is organized as follows. In section~\ref{section2} we introduce the Boltzmann equation for both Maxwell and VHP models of probabilistic ballistic annihilation, as well as the balance equations for the coarse-grained fields. In section~\ref{section3} we briefly describe the Chapman-Enskog scheme while section~\ref{section4}  is devoted to the Maxwell model. We first find the homogeneous state, and solve the corresponding homogeneous balance equations. To first order in the Chapman-Enskog expansion we then study the effect of a small spatial inhomogeneity. We follow the traditional route to compute the transport coefficients, which consists in truncating the first-order velocity distribution function to the first nonzero term in a Sonine polynomial expansion~\cite{coppexdroztrizac}. We then show that this truncation does not constitute an approximation for the transport coefficients since they can be obtained by solving the Maxwell model \emph{exactly} to first order. The VHP model is subsequently investigated in section~\ref{section5}. We first find the homogeneous cooling state, and then solve the corresponding homogeneous equations. We implement Monte Carlo simulations in order to check the decay exponents found analytically. Next, we establish the transport coefficients to first order in the Chapman-Enskog expansion before presenting a comparison of the transport coefficients of the different models. In Sec.~\ref{section6} we finally perform a linear stability analysis of the Navier-Stokes hydrodynamic equations around the spatially homogeneous state, and compare the results with PBA of hard spheres. We show as well that the second order decay rates may accurately be neglected. Our main findings and conclusions are summarized in section~\ref{section7}. 

Since the underlying calculations of this paper are cumbersome, we present only the main steps in order to focus onto the more relevant results. Further technical details or explanations may be found in~\cite{coppexdroztrizac} and for
convenience, Appendix~\ref{appendixnotations} contains a summary of the notations used.

%=========================================================
\section{The Balance Equations}\label{section2}
%=========================================================

The Boltzmann equation for the one particle distribution $f(\mathbf{r},\mathbf{v};t)$ of particles annihilating upon collision with probability $p$ reads
\begin{equation}
(\partial_t + \mathbf{v}_1 \cdot \boldsymbol{\nabla}) f(\mathbf{r},\mathbf{v}_1;t) = p J_a[f,f] + (1-p) J_c[f,f], \label{boltzmann}
\end{equation}
where $J_a$ is the annihilation operator defined by 
\begin{equation}
J_a[f,g] = - \sigma^{d-1} \phi(x) v_T^{1-x} g(\mathbf{r},\mathbf{v}_1;t) \int_{\mathbbm{R}^d} d \mathbf{v}_2 \, v_{12}^x f(\mathbf{r},\mathbf{v}_2;t)\label{defja}
\end{equation}
and $J_c$ is the collision operator:
\begin{equation}
J_c[f,g] = \sigma^{d-1} \frac{\phi(x) v_T^{1-x}}{S_d} \int_{\mathbbm{R}^d} d \mathbf{v}_2 \, v_{12}^x \int d \widehat{\boldsymbol{\sigma}}\, (b^{-1} -1 ) g(\mathbf{r},\mathbf{v}_1;t) f(\mathbf{r},\mathbf{v}_2;t).\label{defjc}
\end{equation}
In these equations, $d$ denotes the spatial dimension, $v_{12} = |\mathbf{v}_1 - \mathbf{v}_2|$ is the modulus of the relative velocity, $S_d = 2 \pi^{d/2} / \Gamma(d/2)$ is the solid angle surface, $\Gamma$ the Euler gamma function, $v_T = \sqrt{2/\beta m}$ the time-dependent thermal velocity, $\beta=(k_B T)^{-1}$, $\sigma$ is the diameter of the particles, $\widehat{\boldsymbol{\sigma}}$ is a unit vector joining the centers of two particles and the corresponding integral is running over the solid angle. Finally, $b^{-1}$ an operator acting on the velocities as follows:
\begin{subequations}
\begin{eqnarray}
b^{-1} \mathbf{v}_{12} &=& \mathbf{v}_{12} - 2 (\mathbf{v}_{12} \cdot \widehat{\boldsymbol{\sigma}}) \widehat{\boldsymbol{\sigma}},\\
b^{-1} \mathbf{v}_{1} &=& \mathbf{v}_{1} - (\mathbf{v}_{12} \cdot \widehat{\boldsymbol{\sigma}}) \widehat{\boldsymbol{\sigma}}.
\end{eqnarray}\label{definitionoftheoperatorb}
\end{subequations}
The choice $x=0$ ($x=2$) corresponds to the Maxwell (VHP) model, respectively. For hard sphere dynamics, that would correspond to $x=1$, the relative velocity $v_{12}$ gives the rate of collision and its presence makes analytical progress difficult. A convenient simplification \cite{matthieu} to overcome this difficulty is to replace it by $v_{12}^{x} v_T^{1-x}$ where $v_T$ is introduced for dimensional reasons. The quantity $\phi(x)$ which sets the relevant time scale in the problem can be freely chosen, and will be used in the following analysis to obtain the desired limiting behaviour in the limit $p\to 0$ (see also \cite{santos} for related considerations). We also note that particles interacting with an inverse power-law potential are described by a kinetic equation of the same form as Eq.~(\ref{defjc})~\cite{matthieu}.

In order to write hydrodynamic equations, we need to define local hydrodynamic fields:
\begin{subequations}
\begin{eqnarray}
n(\mathbf{r},t) &=& \int_{\mathbb{R}^d} d \mathbf{v} \, f(\mathbf{r},\mathbf{v};t), \label{defn} \\
\mathbf{u}(\mathbf{r},t) &=& \frac{1}{n(\mathbf{r},t)} \int_{\mathbb{R}^d} d \mathbf{v} \, \mathbf{v} f(\mathbf{r},\mathbf{v};t),\label{defu} \\
T(\mathbf{r},t) &=& \frac{m}{n(\mathbf{r},t) k_B d} \int_{\mathbb{R}^d} d \mathbf{v} \, \mathbf{V}^2 f(\mathbf{r},\mathbf{v};t), \label{deft}
\end{eqnarray}
\end{subequations}
where $n(\mathbf{r},t)$, $\mathbf{u}(\mathbf{r},t)$, and $T(\mathbf{r},t)$ are the local number density, velocity, and temperature, respectively (the latter definition being kinetic with no thermodynamic basis). The definition of the temperature follows from the principle of equipartition of energy. In Eq.~(\ref{deft}), $k_B$ is the Boltzmann constant and $\mathbf{V} = \mathbf{v} - \mathbf{u}(\mathbf{r},t)$ is the deviation from the mean flow velocity. The balance equations follow from integrating the moments $1$, $m \mathbf{v}$, and $m v^2/2$ with weight given by the Boltzmann equation~(\ref{boltzmann}). Following the same route as in~\cite{coppexdroztrizac} we thus obtain
\begin{subequations}
\begin{eqnarray}
& & \partial_t n + \nabla_i (n u_i) = - p \omega[f,f], \\
& & \partial_t u_i + \frac{1}{m n} \nabla_j P_{ij} + u_j \nabla_j u_i = - p \frac{1}{n} \omega[f,V_i f], \qquad i=1,\ldots,d, \\
& & \partial_t T + u_j \nabla_j T + \frac{2}{n k_B d}( P_{ij} \nabla_i u_j + \nabla_j q_j) = p \frac{T}{n} \omega[f,f] - p \frac{m}{n k_B d} \omega[f,V^2 f],
\end{eqnarray}\label{balanceequations}
\end{subequations}
with implicit summation over repeated indices, $\mathbf{u}=(u_1, \ldots , u_d)$, and 
\begin{equation}
\omega[f,g] = - \int_{\mathbbm{R}^d} d \mathbf{v}_1 \, J_a[f,g].
\end{equation}
In the balance equations~(\ref{balanceequations}), the pressure tensor $P_{ij}$ and heat-flux $q_i$ are defined by
\begin{equation}
P_{ij}(\mathbf{r},t) = m \int_{\mathbb{R}^d} d \mathbf{v} \, V_i V_j f(\mathbf{r},\mathbf{v};t) =  \int_{\mathbb{R}^d} d \mathbf{v} \, f(\mathbf{r},\mathbf{v};t) D_{ij}(\mathbf{V}) + \frac{n}{\beta} \delta_{ij}, \label{pressure}
\end{equation}
\begin{equation}
q_i(\mathbf{r},t) = \int_{\mathbb{R}^d} d \mathbf{v} S_i(\mathbf{V}) f(\mathbf{r},\mathbf{v};t), \label{heat}
\end{equation}
where $\beta=1/(k_B T)$ and 
\begin{equation}
D_{ij}(\mathbf{V}) = m \left( V_i V_j - \frac{V^2}{d} \delta_{ij} \right), \label{definitiondij}
\end{equation}
\begin{equation}
S_i(\mathbf{V}) = \left( \frac{m}{2} V^2 - \frac{d+2}{2} k_B T \right) V_i.\label{definitionsi}
\end{equation}
As expected, when the annihilation probability $p \to 0$, all three coarse-grained fields $n$, $\mathbf{u}$, and $T$ are conserved.

%=========================================================
\section{The Chapman-Enskog solution}\label{section3}
%=========================================================

The Chapman-Enskog method allows from Eqs.~(\ref{balanceequations}) to build a closed set of equations for the hydrodynamic fields (see e.g \cite{duftycondmat,new}). For this purpose, it  is required to express the functional dependence of the pressure tensor $P_{ij}$ and of the heat flux $q_i$ in terms of the hydrodynamic fields. The Chapman-Enskog approach relies on two important assumptions. The first one is the existence of a normal solution in which all temporal and spatial dependence of the distribution function $f(\mathbf{r},\mathbf{v};t)$ may be expressed in terms of the hydrodynamic fields, $f(\mathbf{r},\mathbf{v};t) = f\left[   \mathbf{v}, n(\mathbf{r},t), \mathbf{u}(\mathbf{r},t), T(\mathbf{r},t)\right]$. The discussion of the relevance of this first assumption can be found elsewhere~(e.g in \cite{duftycondmat}). The second assumption is based on the separation of the microscopic time scale (the average collision time between particles) and macroscopic time scale (the evolution of the hydrodynamic fields and their inhomogeneities). This separation implies that the hydrodynamic fields are only weakly inhomogeneous, which allows for a series expansion in the gradients of the fields, $f = f^{(0)} + \varepsilon f^{(1)} + \varepsilon^2 f^{(2)} + \ldots$,  where each power of the formal small parameter $\varepsilon$ is associated to a given order in spatial gradients. The Chapman-Enskog method assumes the existence of an associated time derivative hierarchy: $\partial/\partial t = \partial^{(0)}/\partial t + \varepsilon \partial^{(1)} / \partial t + \varepsilon^2 \partial^{(2)}/\partial t + \ldots$. The insertion of these expansions in the Boltzmann equation yields 
\begin{equation}
\left( \sum_{k \geqslant 0} \varepsilon^k \frac{\partial^{(k)}}{\partial t} + \mathbf{v}_1 \cdot \boldsymbol{\nabla} \right) \sum_{l \geqslant 0} \varepsilon^{l} f^{(l)} = p J_a\left[\sum_{l \geqslant 0} \varepsilon^{l} f^{(l)},\sum_{l \geqslant 0} \varepsilon^{l} f^{(l)} \right] + (1-p) J_c \left[ \sum_{l \geqslant 0} \varepsilon^{l} f^{(l)}, \sum_{l \geqslant 0} \varepsilon^{l} f^{(l)}\right]. \label{expansion}
\end{equation}
The Chapman-Enskog solution is obtained upon solving the equations order by order in $\varepsilon$.

%=========================================================
\section{The Maxwell Model}\label{section4}
%=========================================================

\subsection{The homogeneous state}\label{sechcsmax}
%----------------------------------------- 
To zeroth order in the gradients, Eq.~(\ref{expansion}) gives
\begin{equation}
\partial_t^{(0)} f^{(0)} = p J_a [ f^{(0)},f^{(0)} ] + (1-p) J_c [ f^{(0)},f^{(0)} ]. \label{eqorder0}
\end{equation}
This equation has a solution, describing the homogeneous state, 
and which obeys the scaling relation
\begin{equation}
f^{(0)}(\mathbf{r},\mathbf{v};t) = \frac{n(t)}{v_T^d(t)} \widetilde{f}(c), \label{eqscaling}
\end{equation}
where $v_T = [2/(\beta m)]^{1/2}$ is the time dependent thermal velocity, and $c=V/v_T$, $\mathbf{V} = \mathbf{v} - \mathbf{u}$. The existence of a scaling solution of the form~(\ref{eqscaling}) seems to be a general feature present in different but related contexts ~\cite{trizac,santos,trizacbound}. This solution being isotropic, one has $\mathbf{u}=0$.

Santos and Brey~\cite{santosbrey} showed that there exists a relationship between the homogeneous solutions of the Maxwell model with $p=0$ and $p\neq 0$. We shall here briefly reproduce their arguments. It is possible to rewrite the Boltzmann equation~(\ref{eqorder0}) for $x=0$ under the form
\begin{equation}
\partial_{t'}^{(0)} f^{(0)}(\mathbf{v};t') = - (C_S + C_R) n(t') f^{(0)}(\mathbf{v};t') + \int_{\mathbbm{R}^d} d \mathbf{v}_1 \int d \widehat{\boldsymbol{\sigma}} \, \chi(\widehat{\boldsymbol{\sigma}}) f^{(0)}(\mathbf{v};t') f^{(0)}(\mathbf{v}_1;t'), \label{bolhs1}
\end{equation}
where $t'=(1-p)t$, $C_S = \int d \widehat{\boldsymbol{\sigma}} \chi(\widehat{\boldsymbol{\sigma}})$, $\chi(\widehat{\boldsymbol{\sigma}})= \sigma^{d-1} \phi(x=1) v_T/S_d$, and $C_R = \int d \widehat{\boldsymbol{\sigma}} \chi(\widehat{\boldsymbol{\sigma}}) p/(1-p)$ is the removal collision frequency. Integrating Eq.~(\ref{bolhs1}) over $\mathbf{v}$, the evolution of the number density is governed by $\partial_{t'} n(t') = - C_R n^2(t')$, the solution being $n(t') = n_0 / (1+ n_0 C_R t')$, where $n_0 = n(t'=0)$. If we define $\tau(t') = \int_0^{t'} d s n(s)/n_0$ and $F(\mathbf{v};\tau) = f^{(0)}(\mathbf{v};t') n_0/n(t')$, then $F(\mathbf{v};\tau)$ satisfies the Boltzmann equation without annihilation (i.e., $C_R=0$). $F(\mathbf{v};\tau)$ therefore evolves towards a Maxwellian, and so does $f^{(0)}$: we have $\widetilde{f}(c) = \mathrm{e}^{-c^2}/\pi^{d/2}$.

\subsection{The zeroth-order Chapman-Enskog solution}
%----------------------------------------
Since $f^{(0)}$ is isotropic, to zeroth order the pressure tensor~(\ref{pressure}) becomes $P_{ij}^{(0)} = p^{(0)} \delta_{ij}$, where $p^{(0)} = n k_B T$ is the hydrostatic pressure, and the heat flux~(\ref{heat}) becomes $\mathbf{q}^{(0)} = 0$. The balance equations to zeroth order read
\begin{subequations}
\begin{eqnarray}
\partial_t^{(0)} n &=& - p n \xi_n^{(0)}, \\
\partial_t^{(0)} u_i &=& - p v_T \xi_{u_i}^{(0)}, \\
\partial_t^{(0)} T &=& - p T \xi_T^{(0)}, \label{maxwelleqzeroth}
\end{eqnarray}\label{balanceequationstoorder0}
\end{subequations}
where the decay rates are
\begin{subequations}
\begin{eqnarray}
\xi_n^{(0)} &=& \frac{1}{n} \omega[f^{(0)},f^{(0)}],\\
\xi_{u_i}^{(0)} &=& \frac{1}{n v_T} \omega[f^{(0)}, V_i f^{(0)}], \qquad i=1,\ldots, d \label{decay0u} \\
\xi_T^{(0)} &=& \frac{m}{n k_B T d} \omega[f^{(0)},V^2 f^{(0)}] - \frac{1}{n} \omega[f^{(0)},f^{(0)}].
\end{eqnarray}\label{maxwelldecayzero}
\end{subequations}
For antisymmetry reasons, one sees from Eq.~(\ref{decay0u}) that $\xi_{u_i}^{(0)}=0$. The calculation of $\xi_n^{(0)}$ and $\xi_T^{(0)}$ are straightforward and give $\xi_n^{(0)} = n \sigma^{d-1} \phi^{\mathrm{M}} v_T$ and $\xi_T^{(0)} = 0$. We have written $\phi^{\mathrm{M}}$ for $\phi(x=0)$. The temperature of the Maxwell model is therefore conserved in the scaling regime (time independent thermal velocity $v_T$). In addition, one has
\begin{equation}
n_H(t) = \frac{n_0}{1+p t \xi_n^{(0)}(0)}, \label{HCSboltzman}
\end{equation}
where the subscript $H$ denotes a quantity evaluated in the homogeneous state, and $\xi_n^{(0)}(0)$ is the decay rate for $t=0$. Note that Eq.~(\ref{HCSboltzman}) was already established in Sec.~\ref{sechcsmax}.

\subsection{The first-order Chapman-Enskog solution}\label{secfirstordermax}\label{secexactboltzce}
%----------------------------------------- 
To first order in the gradients, the Boltzmann equation~(\ref{expansion}) reads
\begin{equation}
[ \partial_t^{(0)} + J ] f^{(1)} = - [ \partial_t^{(1)} + \mathbf{v}_1 \cdot \boldsymbol{\nabla} ] f^{(0)}, \label{boltzmann1}
\end{equation}
the operator $J$ being defined by Eqs.~(\ref{defj}) and~(\ref{defl}). The balance equations~(\ref{balanceequations}) to first order become
\begin{subequations}
\begin{eqnarray}
& & \partial_t^{(1)} n + \nabla_i (n u_i) = - p n \xi_n^{(1)}, \\
& & \partial_t^{(1)} u_i + \frac{k_B}{m n} \nabla_i (n T) + u_j \nabla_j u_i = - p v_T \xi_{u_i}^{(1)}, \qquad i=1,\ldots,d, \\
& & \partial_t^{(1)} T + u_i \nabla_i T + \frac{2}{d} T \nabla_i u_i = - p T \xi_T^{(1)}, \label{balanceequations1energy}
\end{eqnarray}\label{balanceequations1}
\end{subequations}
where the decay rates are given by
\begin{subequations}
\begin{eqnarray}
\xi_n^{(1)} &=& \frac{2}{n} \omega[f^{(0)},f^{(1)}], \\
\xi_{u_i}^{(1)} &=& \frac{1}{n v_T} \omega[f^{(0)},V_i f^{(1)}] + \frac{1}{n v_T} \omega[f^{(1)},V_i f^{(0)}], \qquad i=1,\ldots,d, \\
\xi_T^{(1)}&=& - \frac{2}{n} \omega[f^{(0)},f^{(1)}] + \frac{m}{n k_B T d} \omega[f^{(0)}, V^2 f^{(1)}] + \frac{m}{n k_B T d} \omega[f^{(1)},V^2 f^{(0)}].
\end{eqnarray}\label{decay1}
\end{subequations}

By definition $f^{(1)}$ is of first order in the gradients of the hydrodynamic fields; for a low density gas \cite{new}
\begin{equation}
f^{(1)} = \mathcal{A}_i \nabla_i \ln T + \mathcal{B}_i \nabla_i \ln n + \mathcal{C}_{ij} \nabla_j u_i. \label{f1}
\end{equation}
The coefficients $\mathcal{A}_i$, $\mathcal{B}_i$, and $\mathcal{C}_{ij}$ depend on the fields $n$, $\mathbf{V}$, and $T$.

\subsubsection{The approximate first-order Chapman-Enskog solution}
%----------------------------------------- 

The hydrodynamic description of the flow requires the knowledge of transport coefficients, which may be determined from a Sonine polynomial expansion of the first order distribution function. In addition, the pressure tensor may be put in the form
\begin{equation}
P_{ij}(\mathbf{r},t) = p^{(0)} \delta_{ij} - \eta \left(  \nabla_i u_j + \nabla_j u_i - \frac{2}{d} \delta_{ij} \nabla_k u_k \right) - \zeta \delta_{ij} \nabla_k u_k,\label{pressurephenomenological}
\end{equation}
where $p^{(0)}=n k_B T$ is the ideal gas pressure, $\eta$ is the shear viscosity, and $\zeta$ is the bulk viscosity which vanishes for a low density gas~\cite{duftycondmat}. Fourier's linear law for heat conduction is
\begin{equation}
q_i = - \kappa \nabla_i T - \mu \nabla_i n,\label{heatphenomenological}
\end{equation}
where $\kappa$ is the thermal conductivity and $\mu$ a transport coefficient that has no analogue in the elastic case~\cite{premier,Mareschal}.

The identification of Eq.~(\ref{pressurephenomenological}) with Eq.~(\ref{pressure}) using the result of the first order calculation yields
\begin{equation}
P_{ij}^{(1)} = \int_{\mathbb{R}^d} d \mathbf{v} \, D_{ij}(\mathbf{V}) f^{(1)}.\label{pressure1}
\end{equation}
Similarly, the identification of Eq.~(\ref{heatphenomenological}) with Eq.~(\ref{heat}) using the first order calculation leads to
\begin{equation}
q_i^{(1)} = \int_{\mathbb{R}^d} d \mathbf{v} \, S_i(\mathbf{V}) f^{(1)}. \label{heat1}
\end{equation}

The calculation follows the same route as in~\cite{coppexdroztrizac}, and we obtain
\begin{subequations}
\begin{eqnarray}
\eta^* = \frac{\eta}{\eta_0} \phantom{n}  &=& \frac{1}{\nu_\eta^*},\\
\kappa^* = \frac{\kappa}{\kappa_0} \phantom{n}  &=& \frac{d-1}{d} \frac{1}{\nu_\kappa^*}, \\
\mu^* = \frac{n \mu}{T \kappa_0} &=& 0,
\end{eqnarray}\label{transportcoefficientsapprox}
\end{subequations}
where the thermal conductivity $\kappa_0$ and shear viscosity $\eta_0$ coefficients for hard spheres (used here to obtain dimensionless quantities) are given by Eqs.~(\ref{kappa0}) and~(\ref{eta0}), respectively~\cite{ferziger}. The dimensionless coefficients $\nu_\eta^*$ and $\nu_\kappa^*$ are given by
\begin{subequations}
\begin{eqnarray}
\nu_\kappa^* &=& \frac{1}{\nu_0} \frac{\int_{\mathbb{R}^d} d \mathbf{V} \, S_{i}(\mathbf{V}) J \mathcal{A}_{i}}{\int_{\mathbb{R}^d} d \mathbf{V} \, S_{i}(\mathbf{V}) \mathcal{A}_{i}} - p \frac{1}{\nu_0} \frac{\int_{\mathbb{R}^d} d \mathbf{V} \, S_{i}(\mathbf{V}) \Omega \mathcal{A}_{i}}{\int_{\mathbb{R}^d} d \mathbf{V} \, S_{i}(\mathbf{V}) \mathcal{A}_{i}}, \\
\nu_\eta^* &=& \frac{1}{\nu_0} \frac{\int_{\mathbb{R}^d} d \mathbf{V} \, D_{ij}(\mathbf{V}) J \mathcal{C}_{ij}}{\int_{\mathbb{R}^d} d \mathbf{V} \, D_{ij}(\mathbf{V}) \mathcal{C}_{ij}} - p\frac{1}{\nu_0} \frac{\int_{\mathbb{R}^d} d \mathbf{V} \, D_{ij}(\mathbf{V}) \Omega \mathcal{C}_{ij}}{\int_{\mathbb{R}^d} d \mathbf{V} \, D_{ij}(\mathbf{V}) \mathcal{C}_{ij}},
\end{eqnarray}\label{bigintegrals}
\end{subequations} 
where $\nu_0 = p^{(0)}/\eta_0$, with $p^{(0)} = n k_B T$. Note that the above relations are still exact within the Chapman-Enskog expansion. The approximation consists in truncating the function $f^{(1)}$ to the first nonzero term in a Sonine polynomial expansion:
\begin{subequations}
\begin{eqnarray}
\boldsymbol{\mathcal{A}}(\mathbf{V}) &=& a_1 \mathcal{M}(\mathbf{V}) \mathbf{S}(\mathbf{V}), \\
\boldsymbol{\mathcal{B}}(\mathbf{V}) &=& b_1 \mathcal{M}(\mathbf{V}) \mathbf{S}(\mathbf{V}), \\
\boldsymbol{\mathcal{C}}(\mathbf{V}) &=& c_0 \mathcal{M}(\mathbf{V}) \mathbf{D}(\mathbf{V}),
\end{eqnarray}\label{firstorderexpansion}
\end{subequations}
where $a_1$, $b_1$, and $c_0$ are the coefficients of the development, and $\mathcal{M}(\mathbf{V}) = n/(v_T^d \pi^{d/2}) \exp (- V^2/v_T^2)$ is the
Maxwellian in the homogeneous state. This allows one to compute the relations~(\ref{bigintegrals}), and one finds (see Appendix~\ref{appendix1})
\begin{subequations}
\begin{eqnarray}
\nu_\eta^* &=& \phi^{\mathrm{M}} \frac{\sqrt{2} \Gamma(d/2)}{4 \pi^{(d-1)/2}}  \left[ p \frac{d+2}{2} + (1-p)\right] ,\\
\nu_\kappa^* &=& \phi^{\mathrm{M}} \frac{\sqrt{2} \Gamma(d/2)}{4 \pi^{(d-1)/2}}  \left[ \frac{d+2}{2} + (1-p) \frac{d-1}{d} \right].
\end{eqnarray}
\end{subequations}
The parameter $\phi^{\mathrm{M}}$ governing the collision frequency may be freely chosen to allow for a relevant comparison with hard sphere dynamics (see e.g.~\cite{santos}). We choose $\phi$ such that the transport coefficients are normalized to one for $p\to 0$, that is when all collisions are elastic. It is remarkable that for the Maxwell model a single parameter such as $\phi$ is sufficient to ensure normalization of all the transport coefficients (this will not be the case in the VHP approach). This leads to
\begin{equation}
\phi^{\mathrm{M}} = \frac{4\pi^{(d-1)/2}}{\sqrt{2}\Gamma(d/2) }.\label{choixphi}
\end{equation}
The above value turns out to be the same as the one obtained from the elastic limit of the Maxwell model of granular gases~\cite{santos}. In the latter case, $\phi$ was chosen matching the temperature decay rate with that characterizing the homogeneous cooling state of inelastic hard spheres. With the choice~(\ref{choixphi}) the transport coefficients~(\ref{transportcoefficientsapprox}) become
\begin{subequations}
\begin{eqnarray}
\eta^* &=& \frac{1}{p \frac{d+2}{2} + (1-p)}, \\
\kappa^* &=& \frac{1}{p \frac{d(d+2)}{2(d-1)} + (1-p)}, \\
\mu^* &=& 0,
\end{eqnarray}\label{transportcoefficientsendapprox}
\end{subequations}

Following the same route as in~\cite{coppexdroztrizac}, the first-order distribution function~(\ref{f1}) reads
\begin{equation}
f^{(1)}(\mathbf{r},\mathbf{V};t) = - \frac{\beta^3}{n} \mathcal{M}(\mathbf{V}) \left[ \frac{2m}{d+2} S_i(\mathbf{V})  \kappa \nabla_i T  + \frac{\eta}{\beta} D_{ij}(\mathbf{V}) \nabla_j u_i \right]. \label{f1end}
\end{equation}

\subsubsection{The exact first-order Chapman-Enskog solution}\label{sectionfirstcecorrection}
%----------------------------------------- 
By construction of the Chapman-Enskog method, the velocity moments of $f$ are given by those of the local equilibrium distribution $f^{(0)}$. It is then easy to show that the decay rates to first order~(\ref{decay1}) are equal to zero [therefore $\Omega f^{(1)} = 0$, where the operator $\Omega$ is defined in Appendix~\ref{appendix2}]. Proceeding in a similar way as in~\cite{santos}, we obtain in Appendix~\ref{appendix2} the exact transport coefficients for the Maxwell model, i.e. without any approximation on the form of $f^{(1)}$. This may be done by integrating the Boltzmann equation~(\ref{boltzexacta}) over $\mathbf{V}$ with weight $m V_i V_j$ and $m V^2 V_i /2$. With the choice for $\phi^{\mathrm{M}}$ given by Eq.~(\ref{choixphi}), one finds the same transport coefficients as those given by Eqs.~(\ref{transportcoefficientsendapprox}). This means that the truncation of $f^{(1)}$ to its first nonzero term in a Sonine polynomial expansion is a harmless approximation when looking at the transport coefficients (this is a peculiarity of the Maxwell model). In fact, it turns out that the transport coefficients depend only on the first term in the Sonine polynomial expansion of $f^{(1)}$~\cite{chapmancowling}. For example, the heat current~(\ref{heat}) may be rewritten under the form~\cite{chapmancowling,booksantosgarzo}
\begin{equation}
q_i^{(1)} = - \frac{d+2}{2} \frac{n}{m \beta^3} \left( a_1 \nabla_i T + b_1 \nabla_i n \right),
\end{equation}
where the first nonzero coefficients ($a_1$, $b_1$) (that may depend on $n$ and $T$) in the Sonine expansion are defined by Eqs.~(\ref{firstorderexpansion}). Therefore the latter coefficients always give an exact result for the transport coefficients, but the problem at hand is to calculate them exactly. This turns out to be possible within the Maxwell model.

\subsection{Hydrodynamic equations}\label{sec:hydromax}
%----------------------------------------- 
Since the pressure tensor and the heat flux defined by Eqs.~(\ref{pressurephenomenological}) and~(\ref{heatphenomenological}), respectively, are of order $1$ in the gradients, their insertion in the balance equations~(\ref{balanceequations}) yields contributions of order 2. Knowledge of the second order velocity distribution $f^{(2)}$ is therefore required in order to find the correct decay rates that contribute to Navier-Stokes order. However, we show in Sec.~\ref{section2order} that for the Maxwell gas those contributions are equal to zero for any annihilation probability $p$. The corresponding hydrodynamic Navier-Stokes equations are given by
\begin{subequations}
\begin{eqnarray}
& & \partial_t n + \nabla_i (n u_i) = - p n [\xi_n^{(0)} + \xi_n^{(1)}], \\
& & \partial_t u_i + \frac{1}{m n} \nabla_j P_{ij} + u_j \nabla_j u_i = - p v_T [\xi_{u_i}^{(0)} + \xi_{u_i}^{(1)}], \qquad i=1,\ldots,d, \\
& & \partial_t T + u_i \nabla_i T + \frac{2}{n k_B d}( P_{ij} \nabla_i u_j + \nabla_i q_i) = - p T [\xi_T^{(0)} + \xi_T^{(1)}].
\end{eqnarray}\label{balanceequationsend}
\end{subequations}
$P_{ij}$ and $q_j$ are given by Eqs.~(\ref{pressurephenomenological}) with $\zeta=0$, and~(\ref{heatphenomenological}) respectively. The rates $\xi_n^{(1)}$, $\xi_{u_i}^{(1)}$, and $\xi_T^{(1)}$ may be calculated using their definition~(\ref{balanceequations1}) and the distribution~(\ref{f1end})~\cite{coppexdroztrizac}. We find that all decay rates are equal to zero, except
\begin{equation}
\xi_n^{(0)} = \frac{d+2}{2} \nu_0.
\end{equation}
We thus have a closed set of equations for the hydrodynamic fields to the Navier-Stokes order. 

%=========================================================
\section{The VHP model}\label{section5}
%=========================================================

\subsection{The homogeneous cooling state}
%----------------------------------------- 
Integrating the Boltzmann equation~(\ref{boltzmann}) over $\mathbf{V}$ for $x=2$, one obtains
\begin{equation}
\frac{d n}{d t} = - p \omega (t) n,
\end{equation}
where
\begin{equation}
\omega (t) = n(t) v_T(t) \sigma^{d-1} \phi^{\mathrm{VHP}} \langle c_{12}^2 \rangle,
\end{equation}
and $\langle g(\mathbf{c}_1,\mathbf{c}_2) \rangle = \int_{\mathbbm{R}^{2d}} d \mathbf{c}_1 d \mathbf{c}_2 \, g(\mathbf{c}_1,\mathbf{c}_2) \widetilde{f}(c_1) \widetilde{f}(c_2)$ denotes the average of a function $g(\mathbf{c}_1,\mathbf{c}_2)$ in the homogeneous cooling state (HCS). We have written $\phi^{\mathrm{VHP}}$ for $\phi(x=2)$. Following the same route as in~\cite{trizac,coppex} the Boltzmann equation may be rewritten in the form
\begin{equation}
\langle c_{12}^2 \rangle \left[ 1 + \frac{1-\alpha_e}{2}\left( d + c_1 \frac{d}{d c_1} \right) \right]  \widetilde{f}(c_1) = \widetilde{f}(c_1) \int_{\mathbbm{R}^d} d \mathbf{c}_2 \, c_{12}^2 \widetilde{f}(c_2) - \frac{1-p}{p} \frac{1}{S_d} \widetilde{I}[\widetilde{f},\widetilde{f}],\label{boltzmoments1}
\end{equation}
where
\begin{equation}
\alpha_e = \frac{\int_{\mathbbm{R}^{2d}} d \mathbf{c}_1 d \mathbf{c}_2 \int d \widehat{\boldsymbol{\sigma}} \, c_{12}^2 c_1^2 \widetilde{f}(c_1) \widetilde{f}(c_2) }{\left[ \int_{\mathbbm{R}^d} d \mathbf{c} \, c^2 \widetilde{f}(c) \right] \int_{\mathbbm{R}^{2d}} d \mathbf{c}_1 d \mathbf{c}_2 \int d \widehat{\boldsymbol{\sigma}} \,  c_{12}^2 \widetilde{f}(c_1) \widetilde{f}(c_2)} = \frac{\langle c_{12}^2 c_1^2\rangle}{\langle c_1^2 \rangle\langle c_{12}^2 \rangle},\label{alphae}
\end{equation}
and
\begin{equation}
\widetilde{I}[\widetilde{f},\widetilde{f}] = \int_{\mathbbm{R}^{d}} d \mathbf{c}_1  \int d \widehat{\boldsymbol{\sigma}} \, c_{12}^2 (b^{-1} -1) \widetilde{f}(c_1) \widetilde{f}(c_2).
\end{equation}

The limit $c_1 \to 0$ of the Boltzmann equation~(\ref{boltzmoments1}) encodes a useful information for ballistically controlled
dynamics~\cite{Lettre,trizac,physica,coppex}:
\begin{equation}
\langle c_{12}^2 \rangle \left( 1 + d \frac{1-\alpha_e}{2} \right)
\widetilde{f}(0) = \widetilde{f}(0) \langle c^2 \rangle - \frac{1-p}{p} 
\frac{1}{S_d} \lim_{c_1 \to 0} \widetilde{I}[\widetilde{f},\widetilde{f}]. 
\label{boltzmoments0}
\end{equation}
Next, we consider the first nonzero correction to the Maxwellian in a Sonine polynomial expansion of the HCS:
\begin{equation}
\widetilde{f}(c) = \widetilde{\mathcal{M}}(c) \left[ 1+a_2 S_2(c^2)\right], \label{firstsonine}
\end{equation}
where $\widetilde{\mathcal{M}}(c) = \pi^{-d/2} \mathrm{e}^{-c^2}$ is the Maxwellian and $S_2(c^2) = c^4/2 - (d+2) c^2/2 + d(d+2) / 8$ the second Sonine polynomial~\cite{chapmancowling}. Eqs.~(\ref{boltzmoments0}) and~(\ref{alphae}) form a system of two equations for the two unknown $\alpha_e$ and $a_2$. Making use of the relations~(\ref{app3eq5}), it is a straightforward task to compute the limit in the right hand side of Eq.~(\ref{boltzmoments0})~\cite{physica}, which gives
\begin{equation}
\lim_{c_1 \to 0} \widetilde{I}[\widetilde{f},\widetilde{f}] = - a_2 \frac{S_d}{\pi^{d/2}} \frac{d^2(d+2)}{16}. \label{finaleone}
\end{equation}
Using Eq.~(\ref{firstsonine}), one easily obtains from Eq.~(\ref{alphae})
\begin{equation}
\alpha_e = \frac{d+1}{d} + a_2 \frac{d+2}{2d}. \label{finaletwo}
\end{equation}
Note that Eqs.~(\ref{finaleone}) and~(\ref{finaletwo}) are exact relations for which all nonlinear contributions in $a_2$ were kept. However, those nonlinear terms cancel out in each case. Making use of $\langle c_{12}^2 \rangle = d$, the insertion of Eqs.~(\ref{finaleone}) and~(\ref{firstsonine}) in~(\ref{boltzmoments0}) gives
\begin{equation}
\left(1- a_2 \frac{d+2}{2} \right) \left[ 1+a_2 \frac{d(d+2)}{8}\right] = 1 + a_2 \frac{d(d+2)}{8} \frac{1}{p}. \label{eqorder2}
\end{equation}
Eq.~(\ref{eqorder2}) admits two solutions, the first one being $a_2 = 0$ and the second one $a_2 = -2 [d+p(4-d)]/[d(d+2)p]$. The second solution is not physical since it diverges for $p=0$. Therefore $a_2 = 0$ and the HCS of the VHP model within the approximation~(\ref{firstsonine}) is described by the local Maxwellian $\widetilde{\mathcal{M}}(c) = \pi^{-d/2} \mathrm{e}^{-c^2}$. We also note that upon discussing the potential ambiguities resulting from such a linearization scheme in $a_2$ (as done in~\cite{montasan,physica}), the same conclusion is reached.

\subsection{The zeroth-order Chapman-Enskog solution}
%----------------------------------------
Proceeding in a similar way as already described, we obtain a set of equations formally identical to Eqs.~(\ref{balanceequationstoorder0}) and~(\ref{maxwelldecayzero}). The calculation of the decay rates gives $\xi_{u_i}^{(0)}=0$, $\xi_T^{(0)} = n \sigma^{d-1} \phi^{\mathrm{VHP}} v_T$, and $\xi_n^{(0)} = \xi_T^{(0)} d$. The HCS is therefore given by
\begin{subequations}
\begin{eqnarray}
n_H(t) = n_0(1+ p t/t_0)^{-\gamma_n},\label{eq47a} \\
T_H(t) = T_0(1+p t/t_0)^{-\gamma_T},\label{secondeqac231}
\end{eqnarray}\label{HCSvhp}
\end{subequations}
where the decay exponents are $\gamma_n = \xi_n^{(0)}(0) t_0$, $\gamma_T =
\xi_T^{(0)}(0) t_0$, and the relaxation time 
$t_0^{-1} = \xi_n^{(0)}(0) + \xi_T^{(0)}/2$. 
In other words, we have 
\begin{equation}
\gamma_n = \frac{2d}{2d+1}, \qquad \mathrm{and}\qquad \gamma_T = \frac{2}{2d+1}. 
\end{equation}
These quantities do not depend either on $\phi$ nor on the annihilation probability $p$. The former result is an exact property of the dynamics under study (the factor $\phi$ may be absorbed into a rescaling of time $t$, leaving scaling exponents unaffected) while the latter may {\it a priori} be an artifact of the approximations made (it will however be shown below that the $p$ dependence --if any-- is extremely weak). If we define the root-mean-square velocity by $\overline{v} = \sqrt{\langle v^2 \rangle}$, then from the definition~(\ref{deft}) of the temperature $\overline{v}(t) \propto T_H^{1/2}(t)$, and from Eq.~(\ref{secondeqac231}) we have $\overline{v} \sim t^{-\gamma_v}$ for long times, with $\gamma_v = \gamma_T/2$. The decay exponents $\gamma_n$ and $\gamma_v$, as well as the decay exponents for the Maxwell model, agree with the prediction of Krapivsky and Sire~\cite{krapivskysire}, and satisfy the scaling constraint $\gamma_n +\gamma_v = 1$, which essentially expresses the unicity of the relevant time scale in the problem. Moreover, making use of the expression for the decay exponents of PBA of hard spheres $\gamma_n^{\mathrm{S}}$ and $\gamma_v^{\mathrm{S}}$ obtained to linear order in $a_2$ and which are recalled in Appendix~\ref{appendixdecayexponentshs}~\cite{coppexdroztrizac,coppex}, it is easy to verify explicitly that the Maxwell and VHP models provide bounds~\cite{krapivskysire}
\begin{equation}
\frac{2d}{2d+1} < \gamma_n^{\mathrm{S}}(p) < 1, \qquad 0 < \gamma_v^{\mathrm{S}}(p) < \frac{1}{2d+1}, \qquad 
\end{equation}
for all $p \in [0,1]$. We emphasized however that the previous inequality have the status of ``empirical'' observations, and could not be anticipated from rigorous arguments.

\begin{figure}
\begin{center}
\includegraphics[width=0.7\columnwidth]{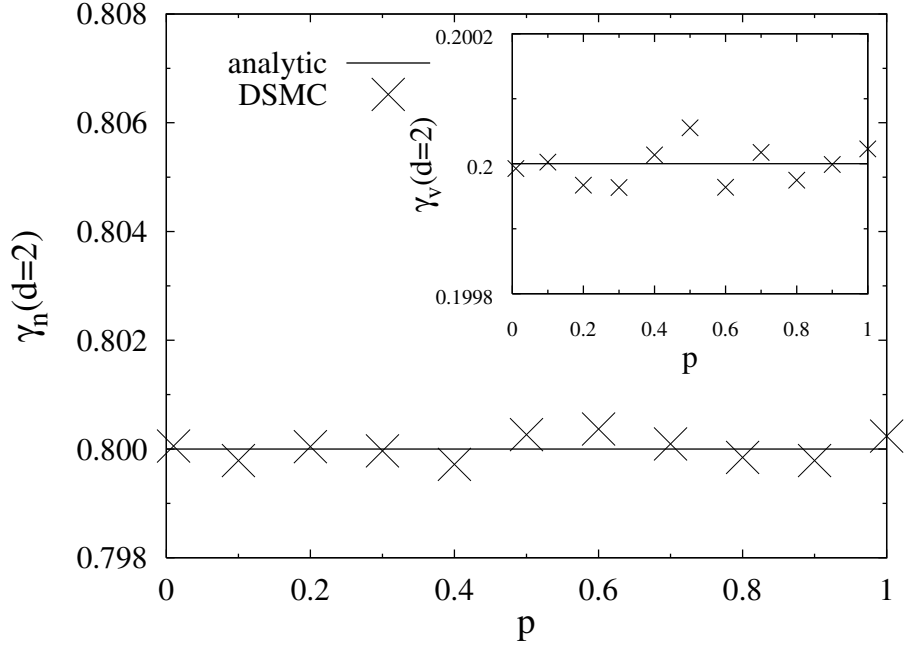}
\end{center}
\caption{The decay exponents $\gamma_n$ and $\gamma_v$ (inset) in two dimensions for the VHP model ($x=2$). The analytical predictions $\gamma_n = 2d/(2d+1) = 0.8$ and $\gamma_v = 1/(2d+1) = 0.2$ are shown by the continuous lines while the symbols stand for the DSMC results (obtained from approximately $300$ independent runs and $10^7$ initial particles). From the above data, it appears that the scaling relation $\gamma_n + \gamma_v = 1$ is well obeyed (the deviation from $1$ does not exceed $4 \times 10^{-4}$) and that the scaling exponents do not depend on $p$. Note the small $y$ scale.}
\label{fig0a}
\end{figure}

\begin{figure}
\begin{center}
\includegraphics[width=0.7\columnwidth]{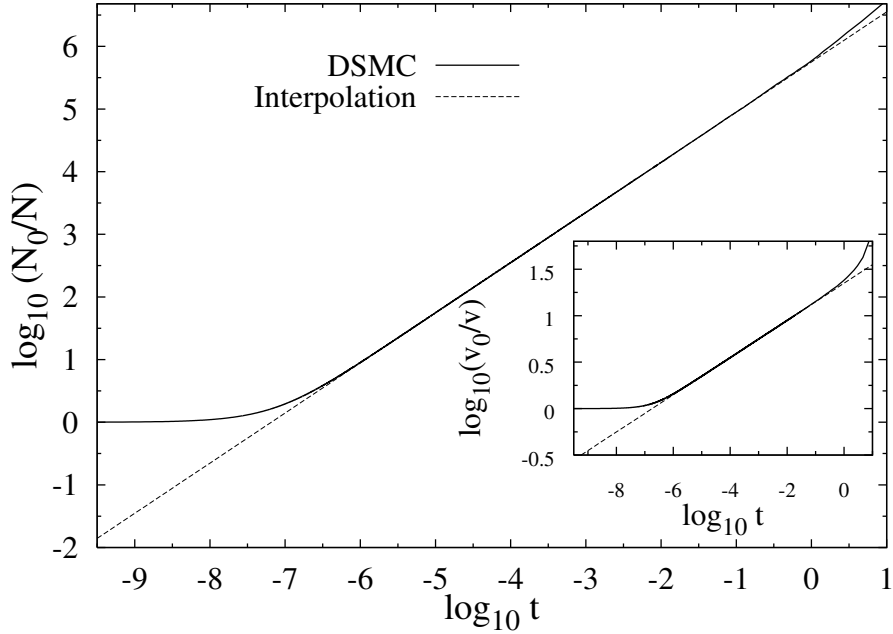}
\end{center}
\caption{Time dependence of $n$ and $\overline{v}$ (inset) for $d=2$ and $p=0.5$ on a log-log scale. The initial velocity distribution is Gaussian. $N_0$ (resp. $N$) is the initial (resp. remaining) number of particles. $v_0 = \overline{v}(0)$ is the root-mean-square velocity at $t=0$, whereas we write $v$ for $\overline{v}(t > 0)$. The dashed straight line is a linear interpolation giving the decay exponent of the power-law, and the deviations to this law for large times is due to the low number of remaining particles.}
\label{fig0b}
\end{figure}

We performed Direct Monte Carlo Simulations (DSMC) in order to verify the decay exponents of the VHP model. The algorithm is similar to the one described in~\cite{trizacbound,coppex}. For the sake of completeness, we briefly outline the main steps of the algorithm. We choose at random two different particles $\{i,j\}$. The time is then increased by $v_T/(N^2 v_{ij}^2)$ where $N$ is the number of remaining particles. With probability $p$ the two particles are removed from the system, and with probability $1-p$ their velocities are modified according to Eqs.~(\ref{definitionoftheoperatorb}). As the fluctuations increase for small $N$, it is necessary to average over several independent realizations in order to diminish the noise. A log-log plot of the density $n/n_0$ and the root-mean-squared velocity $\overline{v}/v_0$ as a function of time gives the decay exponents (see Fig.~\ref{fig0a}). The DSMC results are in excellent agreement with the analytical predictions and the expected power-law behaviors are observed over several decades (see. Fig.~\ref{fig0b}).

\subsection{The approximate first-order Chapman-Enskog solution}
%----------------------------------------- 
The procedure is similar to the one followed within 
the Maxwell model of Sec.~\ref{secfirstordermax} (or~\cite{coppexdroztrizac}), and we find
\begin{subequations}
\begin{eqnarray}
\eta^*&=& \frac{1}{\nu_\eta^* - \frac{1}{2} p \xi_T^{(0)*}},\\
\kappa^* &=& \frac{d-1}{d} \frac{2 \nu_\mu^* - 2 p \xi_n^{(0)*} - 3 p \xi_T^{(0)*}}{X},\\
\mu^* &=& 2 p \frac{d-1}{d}\frac{\xi_T^{(0)*}}{X},
\end{eqnarray}\label{systgde1}
\end{subequations}
where $X=\nu_\kappa^* [2\nu_\mu^*-2p \xi_n^{(0)*}- 3 p \xi_T^{(0)*}] + p \xi_T^{(0)*}\{-4\nu_\mu^* + 3 p[\xi_n^{(0)*}+2\xi_T^{(0)*}]\} $,
\begin{equation}
\nu_\mu^* = \frac{1}{\nu_0} \frac{\int_{\mathbb{R}^d} d \mathbf{V} \, S_{i}(\mathbf{V}) J \mathcal{B}_{i}}{\int_{\mathbb{R}^d} d \mathbf{V} \, S_{i}(\mathbf{V}) \mathcal{B}_{i}} - p \frac{1}{\nu_0} \frac{\int_{\mathbb{R}^d} d \mathbf{V} \, S_{i}(\mathbf{V}) \Omega \mathcal{B}_{i}}{\int_{\mathbb{R}^d} d \mathbf{V} \, S_{i}(\mathbf{V}) \mathcal{B}_{i}}, \label{lastone}
\end{equation}
and $\xi_n^{(0)*}=\xi_n^{(0)}/ \nu_0$, $\xi_T^{(0)*}=\xi_T^{(0)}/\nu_0$. Truncating the function $f^{(1)}$ to the first term in a Sonine polynomial expansion as it was the case for Eqs.~(\ref{firstorderexpansion}), the coefficients $\nu_\eta^*$, $\nu_\kappa^*$, and $\nu_\mu^*$ may be calculated with the help of Appendix~\ref{appendix1}. We find
\begin{subequations}
\begin{eqnarray}
\nu_\eta^* &=& \phi^{\mathrm{VHP}} \frac{\sqrt{2} \Gamma(d/2)}{4 \pi^{(d-1)/2}} \left[ p \frac{(d+2)^2}{2} + (1-p) \frac{(d+2)(d+4)}{4} \right],\\
\nu_\kappa^* = \nu_\mu^* &=&\phi^{\mathrm{VHP}} \frac{\sqrt{2} \Gamma(d/2)}{4 \pi^{(d-1)/2}}  \left[ p \frac{(d+2)(d+3)}{2} + (1-p) \frac{(d-1)(d+4)}{d} \right].
\end{eqnarray}
\end{subequations}

The free parameter $\phi^{\mathrm{VHP}}$ setting the frequency collision has {\it a priori} no reason for being the same as for the Maxwell model. We choose this quantity such that $\eta^*(p=0) = 1$, which means that the shear viscosity for the VHP gas is set for vanishing $p$ to coincide with the shear viscosity $\eta_0$ of hard spheres. This allows for a better comparison of the transport coefficients for the Maxwell, hard sphere, and VHP models. Other choices for $\phi^{\mathrm{VHP}}$ are possible. The condition $\eta^*(0)=1$ leads to
\begin{equation}
\phi^{\mathrm{VHP}} = \phi^{\mathrm{M}} \frac{4}{(d+2)(d+4)},
\end{equation}
so that
\begin{subequations}
\begin{eqnarray}
\xi_n^{(0)*} &=& \frac{2 d}{d+4},\\
\xi_T^{(0)*} &=& \frac{2}{d+4}.
\end{eqnarray}\label{decay0vhp2}
\end{subequations}

The first order distribution function reads
\begin{equation}
f^{(1)}(\mathbf{r},\mathbf{V};t) = - \frac{\beta^3}{n} \mathcal{M}(\mathbf{V}) \left[ \frac{2m}{d+2} S_i(\mathbf{V}) \left( \kappa \nabla_i T + \mu \nabla_i n\right) + \frac{\eta}{\beta} D_{ij}(\mathbf{V}) \nabla_j u_i \right]. \label{f1endvhp} 
\end{equation}
where the transport coefficients are given by Eqs~(\ref{systgde1}).

\subsection{Hydrodynamic equations}
%----------------------------------------- 
The decay rates to first order may be calculated using the definitions~(\ref{balanceequations1}) and the distribution~(\ref{f1endvhp})~\cite{coppexdroztrizac}, which gives
\begin{subequations}
\begin{eqnarray}
\xi_n^{(1)} &=& 0,\\
\xi_{u_i}^{(1)} &=& - v_T \left( \kappa^* \frac{1}{T} \nabla_i T + \mu^* \frac{1}{n} \nabla_i n \right) \xi_u^*,\\
\xi_T^{(1)} &=& 0,
\end{eqnarray}\label{decayvhporder1end}
\end{subequations}
where
\begin{equation}
\xi_u^* = \frac{d^2 (d+2)^2}{8 (d-1)} \phi^{\mathrm{VHP}} \frac{\sqrt{2} \Gamma(d/2)}{4 \pi^{(d-1)/2}}.
\end{equation}
The Navier-Stokes hydrodynamic equations are thus given by Eqs.~(\ref{balanceequationsend}) with the decay rates~(\ref{decay0vhp2}) and~(\ref{decayvhporder1end}). For consistency, second order contributions in the gradients are again needed when evaluating the decay rates. It will be shown in section~\ref{section2order} that they are small corrections to the previous relations.

%=========================================================
\section{Comparison of the transport coefficients}
\begin{figure}
\begin{center}
\psfrag{p}{$p$}
\psfrag{eta}{$\eta^*$}
\psfrag{Legende no 1}{Maxwell}
\psfrag{Legende no 2}{VHP}
\psfrag{Legende no 3}{Hard spheres}
\includegraphics[width=0.7\columnwidth]{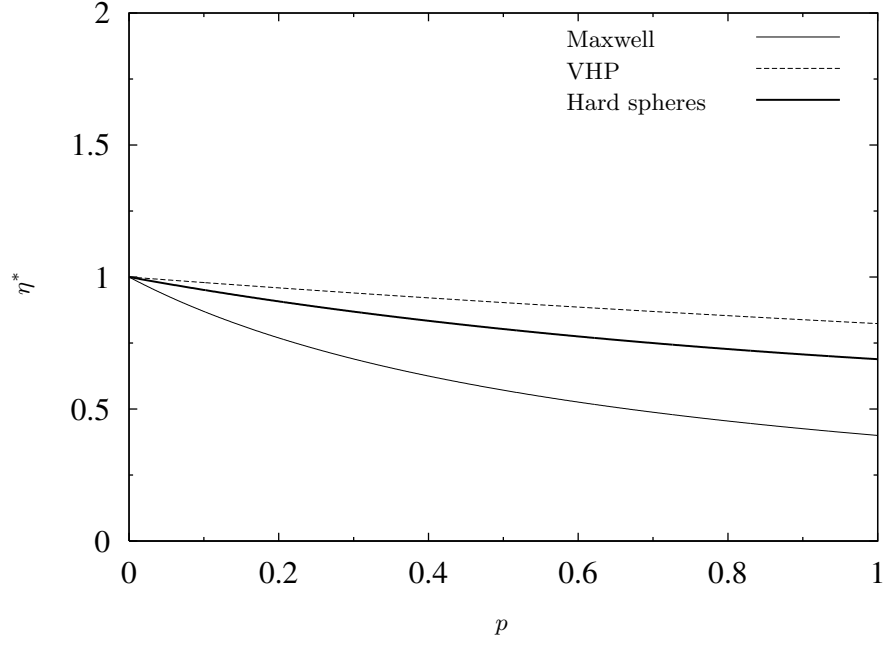}
\end{center}
\caption{Dimensionless shear viscosity $\eta^*$ as a function of the annihilation probability $p$ for the Maxwell (thin continuous line), VHP (dashed line), and hard spheres models (thick continuous line).}
\label{fig1}
\end{figure}

\begin{figure}
\begin{center}
\psfrag{p}{$p$}
\psfrag{kappa}{$\kappa^*$}
\psfrag{Legende no 1}{Maxwell}
\psfrag{Legende no 2}{VHP}
\psfrag{Legende no 3}{Hard spheres}
\includegraphics[width=0.7\columnwidth]{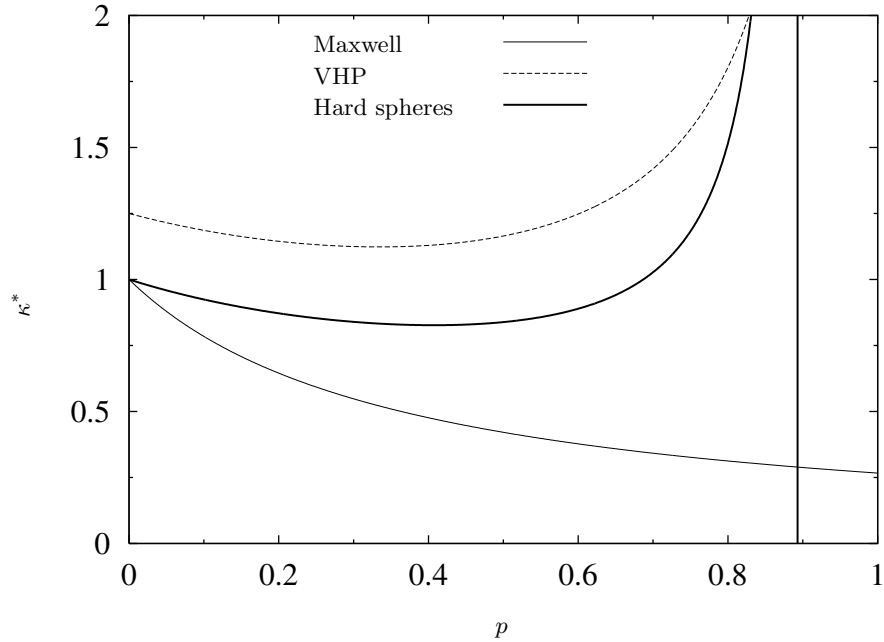}
\end{center}
\caption{Reduced thermal conductivity $\kappa^*$ as a function of the annihilation probability $p$ for the Maxwell (thin continuous line), VHP (dashed line), and hard spheres models (thick continuous line). The vertical lines gives the value $p=0.893\ldots$ for which a divergence of the hard sphere transport coefficients $\kappa^*$ and $\mu^*$  appears (while the shear viscosity exhibits regular behavior, see Fig.~\ref{fig1}).}
\label{fig2}
\end{figure}

\begin{figure}
\begin{center}
\psfrag{p}{$p$}
\psfrag{mu}{$\mu^*$}
\psfrag{Legende no 2}{VHP}
\psfrag{Legende no 3}{Hard spheres}
\includegraphics[width=0.7\columnwidth]{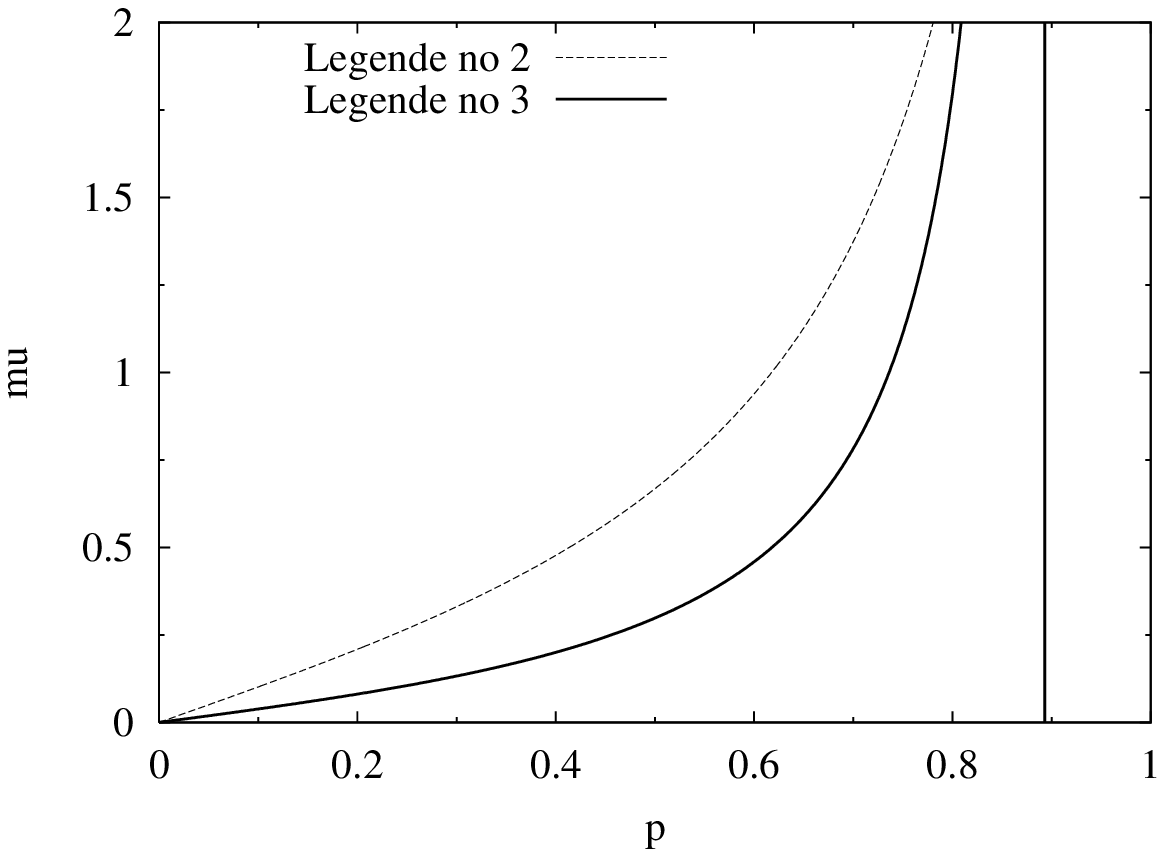}
\end{center}
\caption{Transport coefficient $\mu^*$ as a function of the annihilation probability $p$ (see Fig.~\ref{fig2} for more details). The Maxwell model is not represented since in this case $\mu^* = 0$.}
\label{fig3}
\end{figure}

%=========================================================
We compare the transport coefficients for the Maxwell, VHP, and hard sphere models (the coefficients for the latter model being given in~\cite{coppexdroztrizac}). Figs.~\ref{fig1},~\ref{fig2}, and~\ref{fig3} show $\eta^*$, $\kappa^*$, and $\mu^*$, as a function of the annihilation probability.
Note that once we have chosen $\phi(x=2)$ such that $\eta^* \to 1$ for $p\to 0$ there is no reason to expect $\kappa^*\to 1$ in the same limit. Other choices would have been possible such as enforcing $\kappa^*\to 1$ when $p\to 0$.

From Figures~\ref{fig1},~\ref{fig2}, and~\ref{fig3} it first appears that Mawxell and VHP models capture the essential $p$ dependence of the ``hard sphere'' transport coefficients. In addition, they provide in most cases lower and upper bounds for $\eta^*$, $\kappa^*$ and $\mu^*$. However, as already pointed out in~\cite{coppexdroztrizac}, for strong annihilation probability $p \sim p_d$, the hard sphere thermal conductivity and ``Fourier'' coefficient $\mu$ diverge (see Figs \ref{fig2} and~\ref{fig3}) which leads to a violation of the VHP upper bound for $\kappa$ and $\mu$ in the vicinity of $p_d$. The fact that VHP and Maxwell models lead to smooth and regular transport coefficients for all values of $p$ gives a hint that the hard sphere divergence obtained in previous work \cite{coppexdroztrizac} is a possible artifact of the underlying approximations and probably does not point towards a change of behavior nor a qualitative difference in the scaling or transport properties. This point will be further discussed in the concluding section. We finally note that an {\it a priori} similar deficiency was already reported for the Maxwell model of inelastic hard spheres~\cite{santos}.

%=========================================================
\section{Stability analysis of the Navier-Stokes hydrodynamic equations}\label{section6}
%=========================================================

\subsection{Dispersion relations}\label{section6a}

The hydrodynamic equations.~(\ref{balanceequationsend}) cannot be solved analytically in general. However, their linear stability analysis allows one to answer the question of formation of spatial inhomogeneities. The present study establishes under which conditions the homogeneous state is stable. We consider here a small deviation from spatial homogeneity [see Eqs.~(\ref{HCSboltzman}) and~(\ref{HCSvhp})] and the linearization of Eqs.~(\ref{balanceequationsend}) in the latter perturbation. The procedure used here follows the same route as for granular gases~\cite{premier} or PBA of hard spheres~\cite{coppexdroztrizac}. We define the deviations of the hydrodynamic fields from the homogeneous solution by $\delta y(\mathbf{r},t) = y(\mathbf{r},t) - y_H(t)$, where $y=\{ n, \mathbf{u}, T\}$. Inserting this form in the Navier-Stokes-like equations yields differential equations with time-dependent coefficients. In order to obtain coefficients that do not depend on time, it is necessary to introduce the new dimensionless space and time scales defined by $\mathbf{l} = \nu_{0H}(t) \sqrt{m/[k_B T_H(t)]} \mathbf{r}/2$, $\tau = \int_0^t d s \, \nu_{0H}(s) / 2$, as well as the dimensionless Fourier fields $\rho_\mathbf{k}(\tau) = \delta n_\mathbf{k}(\tau) /n_H(\tau)$, $\mathbf{w}_\mathbf{k}(\tau) = \sqrt{m/[k_B T_H(\tau)]} \delta \mathbf{u}_\mathbf{k}(\tau)$, and $\theta_\mathbf{k}(\tau) = \delta T_\mathbf{k}(\tau) / T_H(\tau)$, where $\delta y_\mathbf{k}(\tau) = \int_{\mathbb{R}^d} d \mathbf{l} \, \mathrm{e}^{-i \mathbf{k} \cdot \mathbf{l}} \delta y(\mathbf{l},\tau)$. Note that $\mathbf{l}$ is defined (up to a constant prefactor) in units of the mean free path for a homogeneous gas of density $n_H(t)$. The dimensionless time $\tau(t)$ gives the accumulated number of collisions per particles up to time $t$. Since we will study both the Maxwell and VHP systems, we recall here the general results valid for non-vanishing decay rates $\xi_n^{(0)}$, $\xi_T^{(0)}$, and $\xi_u^{(1)}$. Making use of the dimensionless variables, the linearized hydrodynamic equations read
\begin{subequations}
\begin{eqnarray}
\left[ \frac{\partial}{\partial \tau} + 2 p \xi_n^{(0)*} \right] \rho_{\mathbf{k}}(\tau) + p \xi_n^{(0)*} \theta_\mathbf{k}(\tau) + i k w_{\mathbf{k}_\parallel}(\tau) =0, \label{stability8a} \\
\left[ \frac{\partial}{\partial \tau} - p \xi_T^{(0)*} + \frac{d-1}{d} \eta^* k^2 \right] \mathbf{w}_{\mathbf{k}_\parallel} + i \mathbf{k} \Big[ \big( 1-p\xi_u^* \mu^* \big) \rho_\mathbf{k}(\tau) + \big( 1-p\xi_u^* \kappa^* \big) \theta_\mathbf{k}(\tau) \Big] = 0, \label{stability8b} \\
\left[ \frac{\partial}{\partial \tau} - p \xi_T^{(0)*} + \frac{1}{2} \eta^* k^2 \right] \mathbf{w}_{\mathbf{k}_\perp}(\tau) = 0, \label{stability8c}\\
\left[ \frac{\partial}{\partial \tau} + p \xi_T^{(0)*} + \frac{d+2}{2(d-1)} \kappa^* k^2 \right] \theta_\mathbf{k}(\tau) + \left[ 2 p \xi_T^{(0)*} + \frac{d+2}{2(d-1)} \mu^* k^2 \right] \rho_\mathbf{k}(\tau) + \frac{2}{d} i k w_{\mathbf{k}_\parallel}(\tau) = 0, \label{stability8d}
\end{eqnarray} \label{stability8}
\end{subequations}
where the transverse mode $\mathbf{w}_{\mathbf{k}_\perp} = \mathbf{w}_{\mathbf{k}} - \mathbf{w}_{\mathbf{k}_\parallel}$ appears to be decoupled from the other equations. The longitudinal velocity field is given by $\mathbf{w}_{\mathbf{k}_\parallel} = (\mathbf{w}_{\mathbf{k}} \cdot \widehat{\mathbf{e}}_\mathbf{k})\widehat{\mathbf{e}}_\mathbf{k}$, and $\widehat{\mathbf{e}}_\mathbf{k}$ is the unit vector along the direction given by $\mathbf{k}$. The transversal velocity field $\mathbf{w}_{\mathbf{k}_\perp}$ consequently defines $(d-1)$ degenerated shear modes. Upon direct integration, we have 
\begin{equation}
\mathbf{w}_{\mathbf{k}_\perp}(\tau) = \mathbf{w}_{\mathbf{k}_\perp}(0) \exp[s_\perp(p,k) \tau], \label{solperp}
\end{equation}
where
\begin{equation}
s_\perp(p,k) = p \xi_T^{(0)*} - \frac{1}{2} \eta^* k^2. \label{modeperp}
\end{equation}
On the other hand, the longitudinal velocity field $\mathbf{w}_{\mathbf{k}_\parallel}$ lies in the one dimensional vector space generated by $\mathbf{k}$. Hence there are three hydrodynamic fields to be determined, namely the density $\rho_\mathbf{k}$, temperature $\theta_\mathbf{k}$, and longitudinal velocity field $\mathbf{w}_{\mathbf{k}_\parallel} = w_{\mathbf{k}_\parallel} \widehat{\mathbf{e}}_\mathbf{k}$. The hydrodynamic matrix $\mathbf{M}$ of the corresponding linear system reads off straightforwardly from Eqs.~(\ref{stability8b})-(\ref{stability8d}). The corresponding eigenmodes are given by $\varphi_n(k) = \exp[s_n(p,k) \tau]$, $n=1,\ldots,3$, where $s_n(p,k)$ are the eigenvalues of $\mathbf{M}$. Each of these three fields is a linear combination of the eigenmodes; thus only the biggest real part of the eigenvalue $s_n(p,k)$ has to be taken into account to discuss the limit of marginal stability of the different modes.

We define $k_\perp$ by the condition $\mathrm{Re}[s_\perp(k_\perp,p)]=0$, i.e.,
\begin{equation}
k_\perp = [2 p \xi_T^{(0)*}/\eta^*]^{1/2},
\end{equation}
and $k_\parallel$ by $\max_{k_\parallel} \mathrm{Re}[s_\parallel(k_\parallel,p)]=0$, $k_\parallel < k_\perp$. Therefore if $k> k_\perp$ all {\em rescaled} modes are linearly stable. For $k \in [k_\parallel,k_\perp]$ only the {\em rescaled} shear mode is linearly unstable (the latter may however be non-linearly coupled to the other modes), and for $k < k_\parallel$ all eigenvalues are positive which leads to instabilities. However, it should be kept in mind that the previous discussion involves rescaled modes only, and should be connected to the original $\mathbf{r}$ variable. Indeed, for any real system (for example a cubic box of volume $L^d$) the smallest wavenumber allowed for a perturbation is given by $2 \pi /L$, which corresponds to the time-dependent dimensionless wavenumber $k_\mathrm{min} = 2 \pi /(L n \sigma^{d-1} {\cal C})$ where  ${\cal C}=4 \pi^{(d-1)/2}/[(d+2)\Gamma(d/2)]$. Since the density $n(t)$ is a decreasing function of time, $k_\mathrm{min}$ increases monotonously and there exists a time $t_\perp$ such that $k_\mathrm{min}(t) > k_\perp$ for $t > t_\perp$. The lower cut-off $k_\mathrm{min}$ therefore eventually enters the region where the homogeneous solution is stable. For $t=t_\perp$, the system is however not in a spatially homogeneous state, but it is nevertheless tempting to conclude that the perturbations will be damped for $t>t_\perp$. Although this statement is not rigorously derived, we conclude here that an instability can only be a transient effect \cite{rque}. 

The time $t_\perp$ can be estimated from the condition $k_\mathrm{min}(t_\perp) = k_\perp$. Making use of the hypothesis of small spatial inhomogeneities, we may replace the density $n(t)$ appearing in the definition of $k_\mathrm{min}(t)$ by the homogeneous density $n_H(t)$ given by Eq.~(\ref{eq47a}). We obtain
\begin{equation}
\frac{t_\perp}{t_0} = \frac{1}{p} \left\{ \left[ \frac{L n_0 \sigma^{d-1} 2 \pi^{(d-3)/2} }{(d+2)\Gamma(d/2)} k_\perp(p) \right]^{1/\gamma_n} -1 \right\}. \label{bluea}
\end{equation}
Is the transient instability alluded to easily observable in a simulation? A typical number of particles for molecular dynamics simulations is of the order of $10^5$, and $n_0 \sigma^2 = 5 \times 10^{-3}$ (which corresponds to a rather low total initial packing function $\pi n_0 \sigma^2 / 4 \simeq 0.004$). For $p=0.1$ and $d=2$ Eq.~(\ref{bluea}) gives $t_\perp \approx 8.6\, t_0\ldots$. Making use of Eq.~(\ref{eq47a}) to approximate the density, one obtains $n(t_\perp) \approx 0.61 n_0$. The density inhomogeneities therefore start to decrease after that the density decreased to only $0.61$ times its initial value, which for $p=0.1$ corresponds in average to only $4$ collisions per particle. 
%The same analysis for PBA of hard spheres~\cite{coppexdroztrizac} gives 
%$t_\perp = 5.1 \,t_0\ldots$, $n(t_\perp)\approx 0.69 n_0$. 
For comparison purposes, inhomogeneities in granular gases are observed after a few hundred collisions per particle~\cite{new,goldetti}. In order to observe the previous (and presumably transient) instabilities one would need molecular dynamics simulations with very large systems. Another condition is to have a large enough $p$, which increases $k_\perp$, see Fig. \ref{fig6}. Equivalently, increasing $p$ increases the divergence rate $s_\perp$ at fixed $k$, see Eq.~(\ref{modeperp}). For sufficiently small $p$ (or small system sizes) Eq.~(\ref{bluea}) does not have a positive solution because $k_\mathrm{min} > k_\perp$ already for $t=0$. To sum up, the typical size of the inhomogeneities may grow as a function of time until $t\simeq t_\perp$ but the subsequent evolution should drive the system back to a time dependent homogeneous regime.

\subsection{The linear second order contributions}\label{section2order}
%----------------------------------------- 
As already pointed out in Sec.~\ref{sec:hydromax}, it is necessary for consistency to include the second order decay rates in the first order Navier-Stokes equations. Their study is useful to establish the relevance of the second order contributions in the gradients [that can be linear in $\nabla^2 n$ or $\nabla^2 T$,
or non-linear in $(\bm{\nabla}T)^2$, $\bm{\nabla}T\cdot\bm{\nabla}n$, etc.]. Our analysis follows the method of Ref.~\cite{premier}, and we shall therefore not enter into much details.

Since we shall perform a linear stability analysis, the only terms in the second order decay rates that will contribute are linear in the gradients. We therefore denote $f_L^{(2)}$ the part of the second order distribution $f^{(2)}$ that yields the linear contributions, and neglect any other term in $f^{(2)}$. The solution has the form $f_L^{(2)} = M(n,T,\mathbf{V}) \nabla^2 T + N(n,T,\mathbf{V}) \nabla^2 n$, where $M$ and $N$ are to be determined. Inserting the latter relation in the linear second order decay rates [that may be obtained from Eqs.~(\ref{decay1}) upon replacing $f^{(1)}$ by $f_L^{(2)}$] they take the form $\xi_A^{(2)} = \xi_{A,1}^{(2)} \nabla^2 T + \xi_{A,2}^{(2)} \nabla^2 n$, where $A=\{n,u_i,T\}$. Again, it is useful to resort to a first order Sonine polynomial expansion for $M$ and $N$, such that $M(\mathbf{V}) = c_T^{(2)} S_2(c^2) \mathcal{M}(\mathbf{V})$ and $N(\mathbf{V}) = c_n^{(2)} S_2(c^2) \mathcal{M}(\mathbf{V})$, where $c_T^{(2)}$ and $c_n^{(2)}$ are the coefficients to be determined.

While it is a rather straightforward task to see from Eqs.~(\ref{decay1}) that those second order contributions to the decay rates are equal to zero for the Maxwell gas, the calculations for the VHP model are more involved. In the latter case all decay rates are equal to zero except $\xi_{T,1}^{(2)} = X c_T^{(2)}$ and $\xi_{T,2}^{(2)} = X c_n^{(2)}$, where $X=\sigma^{d-1} \phi n v_T d(d+2)/2$. The coefficients $c_T^{(2)}$ and $c_n^{(2)}$  can be put in a dimensionless form $c_T^{(2)*} = c_T^{(2)} n k_B T \nu_0 / \kappa_0$ and $c_n^{(2)*} = c_n^{(2)} n^2 k_B \nu_0 /\kappa_0$. Upon replacing the above expansions for $f_L^{(2)}$ and for the decay rates in the Boltzmann equation, the coefficients $c_T^{(2)*}$ and $c_n^{(2)*}$ are solution of the equations
\begin{subequations}
\begin{eqnarray}
\left[ 2 p \xi_n^{(0)*} - p \frac{5}{2} \xi_T^{(0)*} + \nu_\xi^* - p \frac{2(d+2)}{d+4} \right] c_T^{(2)*} &=& p \frac{1}{2} \xi_n^{(0)*} c_n^{(2)*} + \frac{8}{d(d+2)} \kappa^* Y, \\
\left[ p \xi_n^{(0)*} + \nu_\xi^* - p \frac{2(d+2)}{d+4} \right] c_n^{(2)*} &=& p \xi_T^{(0)*} c_T^{(2)*} + \frac{8}{d(d+2)} \mu^* Y,
\end{eqnarray}
\end{subequations}
where $Y= 16 \pi^{(d-1)/2}/[\phi \Gamma(d/2) \sqrt{2} (d+2)(d+4)]$, and
\begin{equation}
\nu_\xi^* = \frac{1}{\nu_0} \frac{\int_{\mathbbm{R}^d} d \mathbf{V} \, V^4 J[S_2(c^2) \mathcal{M}(\mathbf{V})]}{\int_{\mathbbm{R}^d} d \mathbf{V} \, V^4 S_2(c^2) \mathcal{M}(\mathbf{V})} = \phi \frac{\sqrt{2} \Gamma(d/2)}{4\pi^{(d-1)/2}} \left[ p(d+2)(d+3) + (1-p) 4 d \right].
\end{equation}
The second order linear decay rates are given by $\xi_{T,1}^{(2)*} = c_T^{(2)*} d(d+2)/[Y(d-1)(d+4)]$ and $\xi_{T,2}^{(2)*} = c_n^{(2)*} d(d+2)/[Y(d-1)(d+4)]$.

\begin{figure}
\begin{center}
\psfrag{p}{$p$}
\psfrag{this is the y-label}{$\kappa^* /p \xi_{T,1}^{(2)*}$, $\mu^* /p \xi_{T,2}^{(2)*}$}
\psfrag{title 1111}{$\kappa^*/p \xi_{T,1}^{(2)*}$}
\psfrag{title 2222}{$\mu^*/p \xi_{T,2}^{(2)*}$}
\includegraphics[width=0.7\columnwidth]{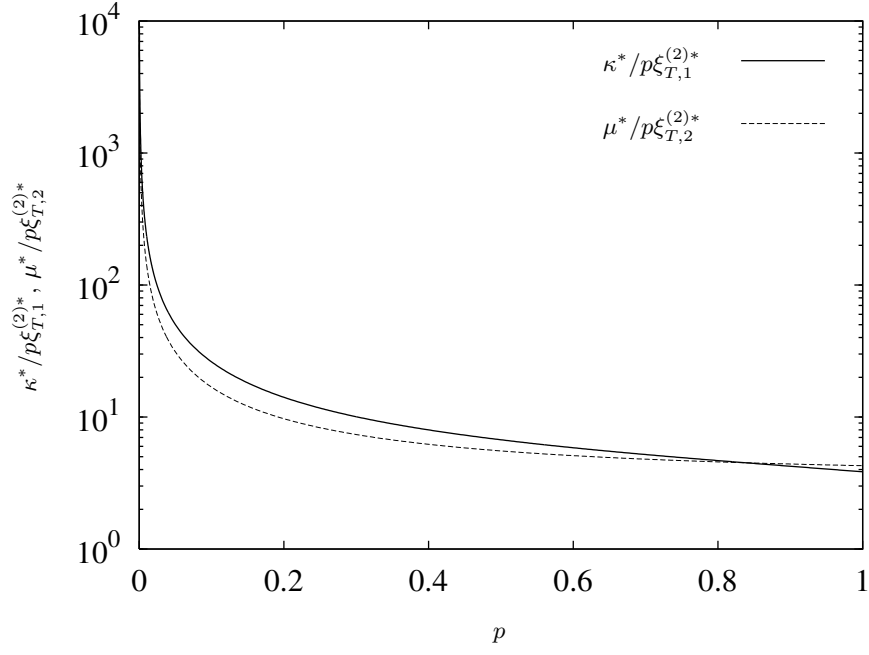}
\end{center}
\caption{Ratio of the dimensionless transport coefficient $\kappa^*$ ($\mu^*$) to the linear second order dimensionless transport coefficients $\xi_{T,1}^{(2)*}$ ($\xi_{T,1}^{(2)*}$), for the VHP model. These ratios do not depend on the the collision frequency $\phi$.}
\label{fig0c}
\end{figure}

The dimensionless linearized hydrodynamic equations with nonzero second order decay rates are the same as Eqs.~(\ref{stability8}) except for Eq.~(\ref{stability8c}) where one has to replace $\kappa^*$ by $\kappa^* - p \xi_{T,1}^{(2)*}$ and $\mu^*$ by $\mu^* - p \xi_{T,2}^{(2)*}$. Consequently, the ratios $\kappa^*/p \xi_{T,1}^{(2)*}$ and $\mu^*/p \xi_{T,2}^{(2)*}$ (see Fig.~\ref{fig0c}) give a rough estimation of the relevance of the second order decay rates.

As seen from Fig.~\ref{fig0c}, the second order decay rates $p \xi_{T,1}^{(2)*}$ and $p \xi_{T,2}^{(2)*}$ are much smaller than the transport coefficients $\kappa^*$ and $\mu^*$, respectively. They may therefore be neglected at least from a linear stability analysis point of view. This approximation is increasingly more accurate as the annihilation probability is decreased. Note that our conclusions for the Maxwell model of probabilistic ballistic annihilation are comparable to those for the granular gas~\cite{premier}.

\subsection{Comparison between Maxwell, very hard particles and hard sphere results}
%----------------------------------------- 

\begin{figure}
\begin{center}
\psfrag{sp1}{$s_\perp$}
\psfrag{sp2}{$s_\parallel$}
\includegraphics[width=0.7\columnwidth]{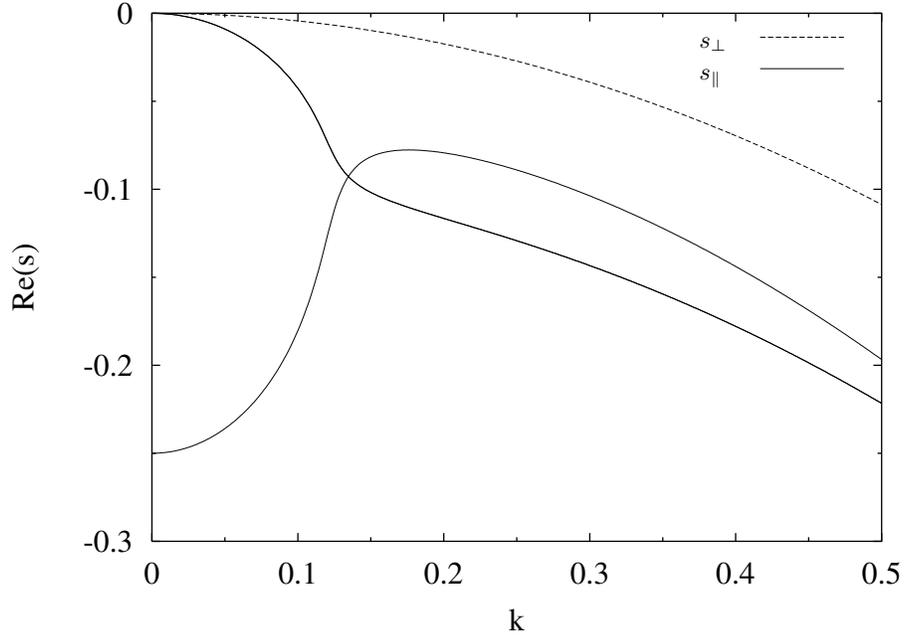}
\end{center}
\caption{Real part of the eigenvalues in dimensionless units for the Maxwell model with $p=0.1$ and $d=3$. The dispersion relation obtained from Eq.~(\ref{modeperp}) is represented by a dashed line (labeled $s_\perp$) whereas the three remaining relations are represented by continuous lines (labeled $s_\parallel$). The shear mode ($s_\perp$) and sound modes (which are on this figure such that $s=0$ when $k \to 0$) are degenerated twice.}
\label{fig4}
\end{figure}

\begin{figure}
\begin{center}
\psfrag{sp1}{$s_\perp$}
\psfrag{sp2}{$s_\parallel$}
\includegraphics[width=0.7\columnwidth]{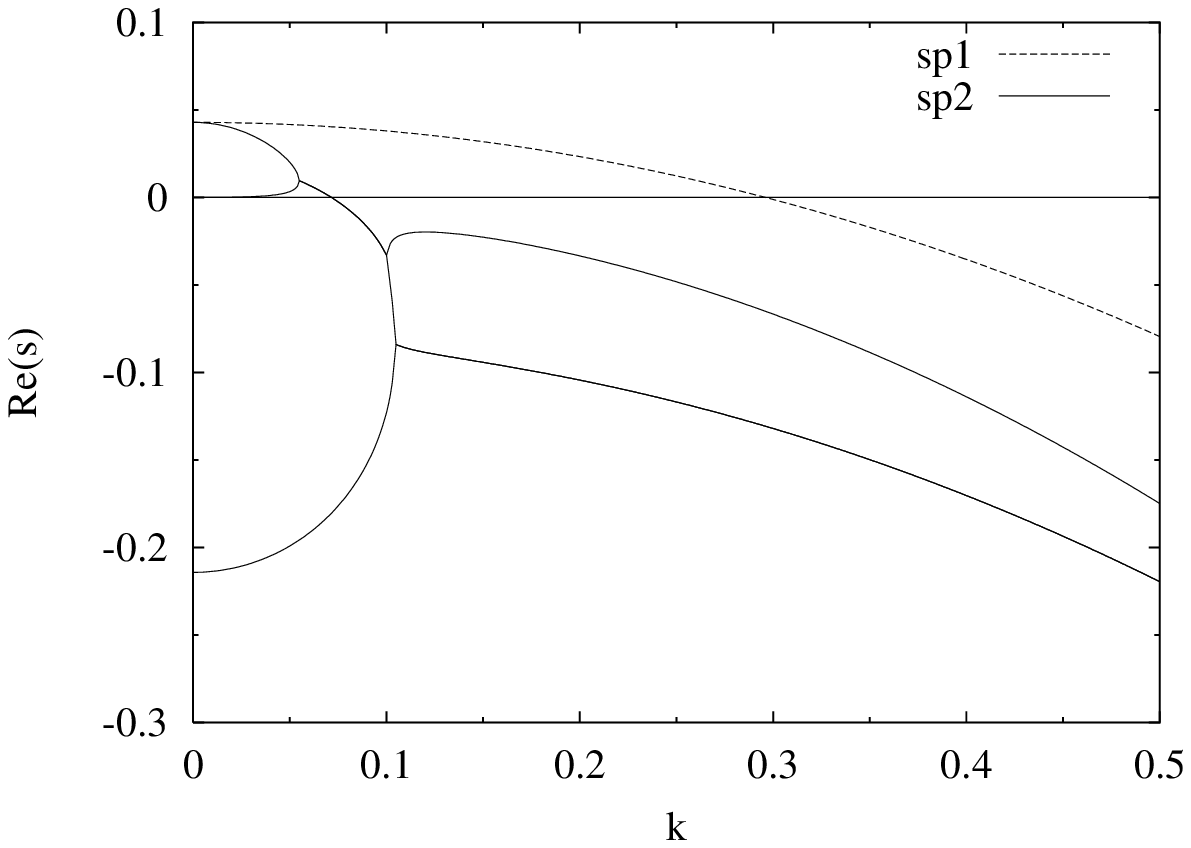}
\end{center}
\caption{Real part of the eigenvalues in dimensionless units for the VHP model with $p=0.1$ and $d=3$. The dispersion relation obtained from Eq.~(\ref{modeperp}) is represented by a dashed line (labeled $s_\perp$) whereas the three remaining relations are represented by continuous lines (labeled $s_\parallel$). The first two biggest parallel modes are sound modes.}
\label{fig5}
\end{figure}

\begin{figure}
\begin{center}
\psfrag{x}{$k_\perp, k_\parallel$}
\psfrag{p}{$p$}
\psfrag{kp1sd}{$k_\perp^{\mathrm{S}}$}
\psfrag{kp2sd}{$k_\parallel^{\mathrm{S}}$}
\psfrag{kp1vhp}{$k_\perp^{\mathrm{VHP}}$}
\psfrag{kp2vhp}{$k_\parallel^{\mathrm{VHP}}$}
\includegraphics[width=0.7\columnwidth]{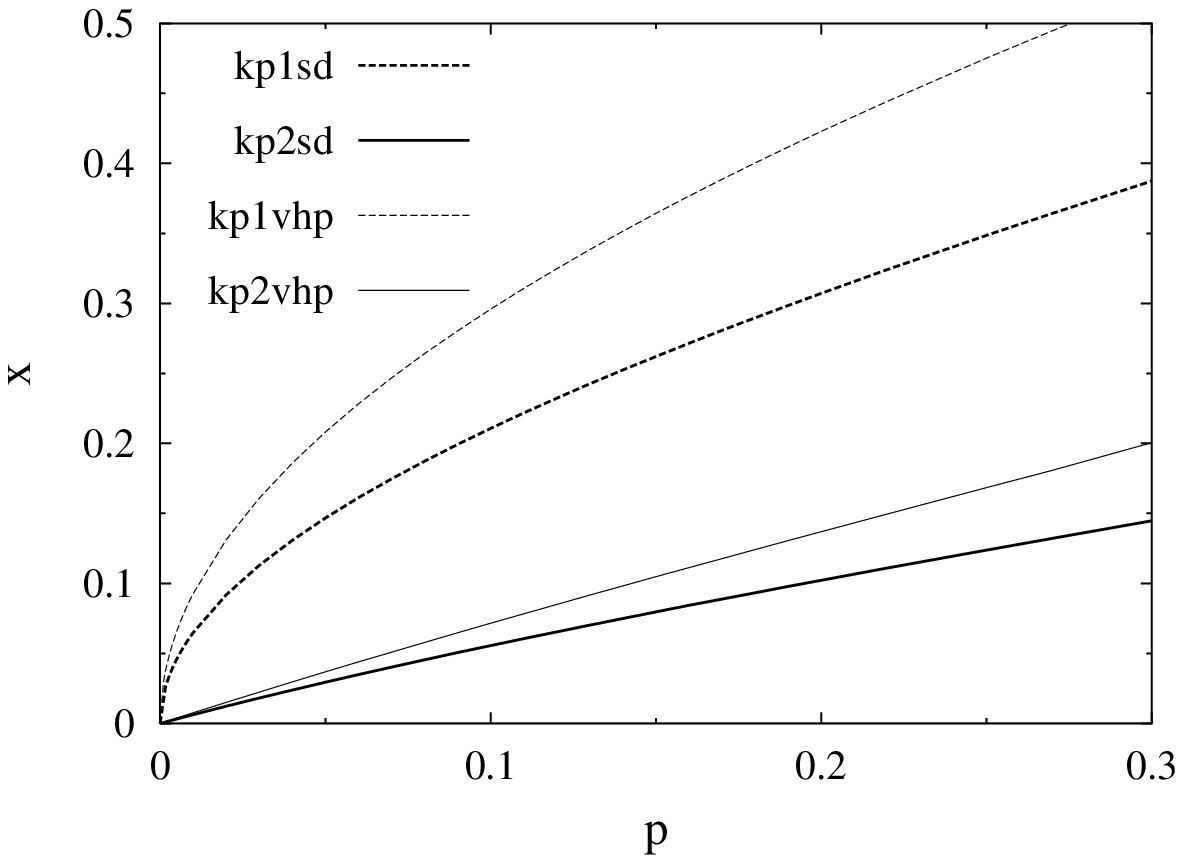}
\end{center}
\caption{Wavenumber $k_\perp$ and $k_\parallel$ in dimensionless units as a function of the annihilation probability $p$ for $d=3$. S and VHP superscripts denote the hard spheres and very hard particles models, respectively. Within the Maxwell model, one has $k_\perp=k_\parallel=0$.}
\label{fig6}
\end{figure}

For the Maxwell model, the temperature decay rate $\xi_T^{(0)}$ vanishes. It follows from Eq.~(\ref{modeperp}) that $k_\perp = 0$ and the transverse mode is stable, which is confirmed by Fig.~\ref{fig4}. The Maxwell model appears to be linearly stable for all values of the annihilation probability $p$.
On the other hand, within the VHP approach, the decay rate $\xi_T{(0)} \neq 0$. The transverse mode may consequently be unstable for some wave-numbers $k$ of the perturbation (see Fig.~\ref{fig5}), which by nonlinear coupling to the other modes may lead to density inhomogeneities. Other modes than the shear may also be linearly unstable, when rescaled wave numbers are such that $k>k_\parallel$. The thresholds $k_\perp$ and $k_\parallel$ are shown in Fig.~\ref{fig6} for the 3 models. It appears again that the hard sphere quantity is bounded below by its Maxwell counterpart and above by VHP. Note that the linear stability analysis does not suffer from arbitrariness related to the free parameter $\phi(x)$. 

\begin{figure}
\begin{center}
\psfrag{sp1}{$s_\perp$}
\psfrag{sp2}{$s_\parallel$}
\psfrag{x}{$k_p$}
\psfrag{p}{$p$}
\psfrag{Im(s)}{\footnotesize $\mathrm{Im}(s_\parallel)$}
\includegraphics[width=0.7\columnwidth]{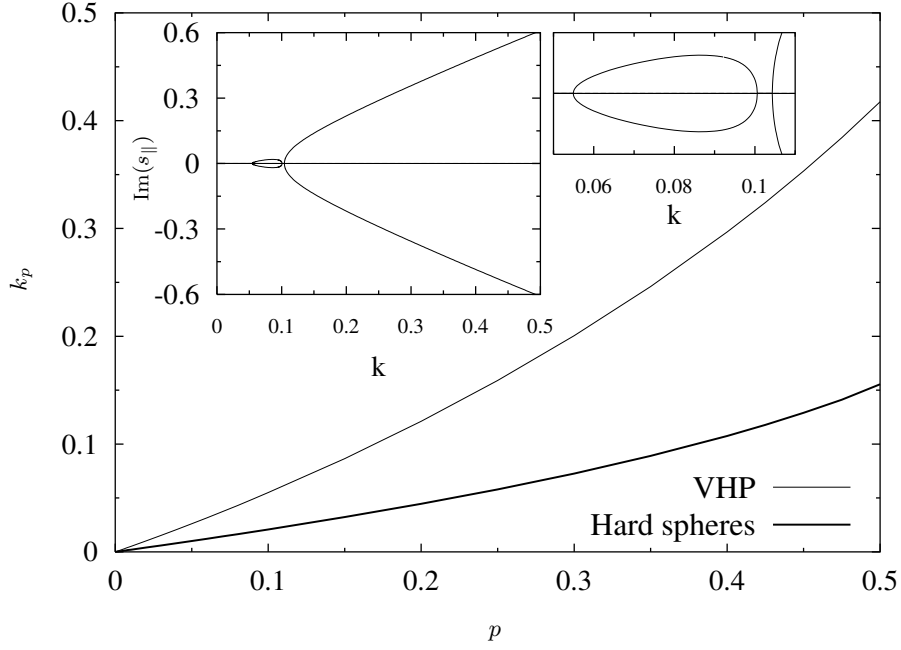}
\end{center}
\caption{Wave number $k_p$ in dimensionless units as a function of the annihilation probability $p$ for $d=3$. The Maxwell model is not represented since in this case $k_p = 0$ for all $p$. The main inset shows the imaginary part of the eigenvalues in dimensionless units for the VHP model for $d=3$ and $p=0.1$. The smaller inset shows the propagation gap $k \in [0.1006\ldots,0.1046\ldots]$ of the sound modes.}
\label{fig7}
\end{figure}

The imaginary part of the eigenvalues embodies the information on the propagation of the perturbations. In Fig.~\ref{fig5}, we identify 3 different parallel modes for small enough $k$ ($k <0.05$). Given that the shear mode is always $(d-1)$ times degenerated and that there are $d+2$ modes in total, none of the parallel modes are degenerated for low enough $k$. Increasing $k$ up to the first bifurcation, the sound modes become degenerated and have a nonzero imaginary value. The non-propagating sound modes thus have bifurcated into a pair of propagating modes. Since the eigenvalue for the transverse velocity field is always real, we shall study here only the imaginary part of the other eigenvalues. We define $k_p$ such that for all $k < k_p$ all eigenvalues are real. It means that only perturbations with small enough wave numbers $\lambda$ such that $\lambda^{-1} > k_p/(2 \pi n \sigma^{d-1})$ are propagating. Fig.~\ref{fig7} shows $k_p$ as a function of the annihilation probability $p$ for the VHP, hard sphere, and Maxwell models. Once more, the VHP and Maxwell models appear as upper and lower bounds, respectively. 
%Since $k_p(p=0)=0$, propagation is only possible if there is a nonzero 
%annihilation probability, what is already known from 
%the transport coefficients. 
From Fig.~\ref{fig4} the Maxwell sound modes are degenerated for all $k$ and therefore the sound modes of the Maxwell model are always propagating, i.e., $k_p = 0$. In the VHP case, Fig. \ref{fig5} shows a propagation gap for the sound modes, i.e., a $k$- window with $k > k_p$, where the sound modes are not degenerated. This is confirmed by Fig.~\ref{fig7} (smaller inset). A propagation gap in the sound mode dispersion relation has been predicted on the basis of the revised Enskog theory for hard-sphere fluids~\cite{reference1}. Such a gap has been observed in neutron scattering experiments of atomic liquids~\cite{reference2}, of molten salts~\cite{reference3}, or inelastic x-ray scattering of lipid bilayers~\cite{reference4} for example.

%=========================================================
\section{Conclusion}\label{section7}
%=========================================================

Making use of the Chapman-Enskog scheme, we have derived in this paper the hydrodynamic equations governing the coarse-grained number density, linear momentum and kinetic energy density fields for an assembly of particles undergoing annihilating collisions with probability $p$ and an elastic scattering otherwise. In between collisions, the motion is ballistic. To this aim, the relevant ``hard sphere''-like Boltzmann equation has been simplified first into its Maxwell, and second into its very hard particle (VHP) form. In both cases, the corresponding Navier-Stokes equations take the same form as in the initial hard sphere description and read
\begin{subequations}
\begin{eqnarray}
& & \partial_t n + \boldsymbol{\nabla} \cdot (n \mathbf{u}) = - p n \, \xi_n, \\
& & \partial_t \mathbf{u} + \frac{1}{m n} \boldsymbol{\nabla} \cdot \mathbf{P} + (\mathbf{u} \cdot \boldsymbol{\nabla}) \mathbf{u} = - p v_T\, \boldsymbol{\xi}_{\mathbf{u}},\\
& & \partial_t T + (\mathbf{u}\cdot \boldsymbol{\nabla}) T + \frac{2}{n k_B d}( \mathbf{P}: \boldsymbol{\nabla} \mathbf{u} + \boldsymbol{\nabla}\cdot \mathbf{q}) = - p T \, \xi_T,
\end{eqnarray}
\end{subequations}
with
\begin{eqnarray}
\xi_n &=& \frac{d+2}{2}\,\nu_0 = 4\, \frac{\pi^{(d-1)/2}}{\Gamma(d/2)} n \sigma^{d-1} \sqrt{\frac{k_B T}{m}},\\
\boldsymbol{\xi}	_{\mathbf{u}}&=& \mathbf{0},\\
\xi_T &=& 0,
\end{eqnarray}
for the Maxwell model, and
\begin{eqnarray}
\xi_n &=& \frac{2d}{d+4}\,\nu_0, \\
\boldsymbol{\xi}_{\mathbf{u}}&=& - v_T \left( \kappa^* \frac{1}{T} \boldsymbol{\nabla} T + \mu^* \frac{1}{n} \boldsymbol{\nabla} n \right) \frac{d^2 (d+2)}{2 (d-1) (d+4)}, \\
\xi_T &=&  \frac{2}{d+4}\,\nu_0,
\end{eqnarray}
in the VHP case [the transport coefficients $\kappa^*$ or $\mu^*$ are given by Eqs.~(\ref{systgde1})].

Our analysis showed that the Maxwell and VHP simplifications, that are more amenable to analytic treatment, not only capture the essential features of hard sphere dynamics, but also provide lower and upper bounds for all comparable quantities. Some important differences should however be commented upon. A first difference is that Maxwell and VHP lead to regular transport coefficients for all values of the annihilation probability, whereas a divergence occurs for annihilating hard sphere thermal conductivity $\kappa$ and Fourier coefficient $\mu$. We concluded from this comparison that this divergence is presumably not physical and could result from the more stringent approximations put forward in the hard sphere computation. It turns out that the hard sphere case is such that the velocity distribution is non-Gaussian to zeroth order in spatial gradient, whereas it is Gaussian in Maxwell and VHP cases. This fact could be at the root of the divergence observed in the transport coefficients. We have also shown that the second order decay rates of the Maxwell model are equal to zero, while for the VHP case they may accurately be neglected. This analysis suggests that the second order decay rates of probabilistic ballistic annihilation of hard spheres may as well be accurately neglected, at least for small annihilation probabilities $p$. This approximation had been invoked in Ref. \cite{coppexdroztrizac}, without any control on its
validity.

The second important difference between Maxwell, hard sphere and VHP dynamics is that within the Maxwell model, all Fourier modes are found to be linearly stable. This fact is intimately related to the non dissipative nature of the corresponding dynamics, an aspect which may be surprising at first: although particles are permanently removed from the system, the mean kinetic energy is conserved on average ($\xi_T=0$). This may be considered as a deficiency of the Maxwell (over)simplification. On the other hand, VHP dynamics is such that the collision frequency increases with the velocity of a given population of particles, which in turn implies that the kinetic energy decreases faster than the number of particles, hence $\xi_T>0$. This dissipation is at the root of possible instabilities in the coarse-grained fields. However, these instabilities manifest themselves for suitably rescaled fields, and we argued in section~\ref{section6} that they should presumably only translate into transient instabilities for the ``real'' fields. Indeed, due to the decrease of density $n(t)$, an unstable Fourier mode has a wavenumber increasing like $n^{-1}$, and eventually enters into a regime where damping should wash out the perturbation. This feature presumably provides at least a linear saturation mechanism for instabilities, different from usual non-linear saturation effects, that may also play a {\em transient} role here if the initial conditions are sufficiently unstable [in other words, if $n(t_\perp)\ll n(t=0)$]. Our stability analysis was nevertheless restricted to perturbations around the time dependent homogeneous state, so that strictly speaking, the effects of transient instabilities that may drive the system into a strongly modulated state are unclear at the moment. This calls for a careful numerical (molecular dynamics) study of the coarse-grained fields, which is the purpose of future work. This would also allow to question the validity of the hydrodynamic description, in a regime where the wave number is not much smaller than the inverse mean-free-path $\ell^{-1} \propto n \sigma^{d-1}$ (in the previous Figures, $k$ is expressed in units of $\ell^{-1}$, up to a prefactor of order one).

%=========================================================
\begin{acknowledgments} 
We acknowledge useful discussions with A. Barrat, A. Puglisi, F. van Wijland, P. Visco, A. Santos, V. Garz\'o, J.J. Brey, M.J. Ruiz-Montero, J. Piasecki, and I. Bena. This work was partially supported by the Swiss National Science Foundation and the European Community's Human Potential Program under contract HPRN-CT-2002-00307 (DYGLAGEMEN).
\end{acknowledgments}
%=========================================================

\appendix
%=========================================================
\section{Summary of the notations}\label{appendixnotations}
%=========================================================
We shall recall here some of the notations used through the paper. $\kappa$ and $\mu$ are the transport coefficients appearing in Fourier's linear heat conduction law~(\ref{heatphenomenological}), and $\eta$ is the shear viscosity appearing in the pressure tensor~(\ref{pressurephenomenological}). A quantity $A$ that is made dimensionless is noted $A^*$. The corresponding dimensionless transport coefficients are written
\begin{subequations}
\begin{eqnarray}
\eta^* &=& \frac{\eta}{\eta_0}, \\
\kappa^* &=& \frac{\kappa}{\kappa_0} \phantom{n}, \\
\mu^* &=& \frac{n \mu}{T \kappa_0},
\end{eqnarray}
\end{subequations}
where
\begin{equation}
\kappa_0 = \frac{d(d+2)}{2(d-1)}\frac{k_B}{m} \eta_0, \label{kappa0}
\end{equation}
\begin{equation}
\eta_0 = \frac{d+2}{8} \frac{\Gamma(d/2)}{\pi^{(d-1)/2}} \frac{\sqrt{m k_B T}}{\sigma^{d-1}}, \label{eta0}
\end{equation}
are the thermal conductivity and shear viscosity coefficients for hard spheres, respectively. The dimensionless coefficients $\nu_\eta^*$, $\nu_\kappa^*$, and $\nu_\mu^*$ are given by
\begin{subequations}
\begin{eqnarray}
\nu_\kappa^* &=& \frac{1}{\nu_0} \frac{\int_{\mathbb{R}^d} d \mathbf{V} \, S_{i}(\mathbf{V}) J \mathcal{A}_{i}}{\int_{\mathbb{R}^d} d \mathbf{V} \, S_{i}(\mathbf{V}) \mathcal{A}_{i}} - p \frac{1}{\nu_0} \frac{\int_{\mathbb{R}^d} d \mathbf{V} \, S_{i}(\mathbf{V}) \Omega \mathcal{A}_{i}}{\int_{\mathbb{R}^d} d \mathbf{V} \, S_{i}(\mathbf{V}) \mathcal{A}_{i}}, \\
\nu_\mu^* &=& \frac{1}{\nu_0} \frac{\int_{\mathbb{R}^d} d \mathbf{V} \, S_{i}(\mathbf{V}) J \mathcal{B}_{i}}{\int_{\mathbb{R}^d} d \mathbf{V} \, S_{i}(\mathbf{V}) \mathcal{B}_{i}} - p \frac{1}{\nu_0} \frac{\int_{\mathbb{R}^d} d \mathbf{V} \, S_{i}(\mathbf{V}) \Omega \mathcal{B}_{i}}{\int_{\mathbb{R}^d} d \mathbf{V} \, S_{i}(\mathbf{V}) \mathcal{B}_{i}}, \\
\nu_\eta^* &=& \frac{1}{\nu_0} \frac{\int_{\mathbb{R}^d} d \mathbf{V} \, D_{ij}(\mathbf{V}) J \mathcal{C}_{ij}}{\int_{\mathbb{R}^d} d \mathbf{V} \, D_{ij}(\mathbf{V}) \mathcal{C}_{ij}} - p\frac{1}{\nu_0} \frac{\int_{\mathbb{R}^d} d \mathbf{V} \, D_{ij}(\mathbf{V}) \Omega \mathcal{C}_{ij}}{\int_{\mathbb{R}^d} d \mathbf{V} \, D_{ij}(\mathbf{V}) \mathcal{C}_{ij}},
\end{eqnarray}\label{bigintegralsappendix}
\end{subequations}
with
\begin{equation}
\nu_0 = \frac{p^{(0)}}{\eta_0} = \frac{8}{d+2}\frac{\pi^{(d-1)/2}}{\Gamma(d/2)} n \sigma^{d-1} \sqrt{\frac{k_B T}{m}},
\end{equation}
and $p^{(0)} = n k_B T$ is the zeroth order pressure. In Eqs.~(\ref{bigintegralsappendix}), the operator $J$ is given by
\begin{equation}
J g = p L_a[f^{(0)},g] + (1-p) L_c[f^{(0)},g],\label{defj}
\end{equation}
where
\begin{subequations}
\begin{eqnarray}
L_a[f^{(0)},g] &=& -J_a[f^{(0)},g] - J_a[g,f^{(0)}],\\
L_c[f^{(0)},g] &=& -J_c[f^{(0)},g] - J_c[g,f^{(0)}],\\
\end{eqnarray}\label{defl}
\end{subequations}
$g$ being an arbitrary function. The collision operator $J_c$ (annihilation operator $J_a$) is defined by Eq.~(\ref{defjc}) [Eq.~(\ref{defja})]. The linear operator $\Omega$ is defined by Eq.~(\ref{defOmega}).

The velocity distribution function is denoted $f(\mathbf{r},\mathbf{v};t)$. In the scaling regime
\begin{equation}
f(\mathbf{r},\mathbf{v};t) = \frac{n(t)}{v_T^d(t)} \widetilde{f}(c),\label{scalingf}
\end{equation}
where $c=V/v_T$. The time dependent [through $T(t)$] thermal velocity is
\begin{equation}
v_T = \sqrt{\frac{2 k_B T}{m}}.
\end{equation}
We note the Maxwellian in the homogeneous cooling state by
\begin{equation}
\mathcal{M}(V) = \frac{n(t)}{v_T^d(t) \pi^{d/2}}\exp \left(-\frac{V^2}{v_T^2}\right),
\end{equation}
and the Maxwellian by
\begin{equation}
\widetilde{\mathcal{M}}(c) = \pi^{-d/2} \exp(-c^2).
\end{equation}
Therefore, we obtain a similar relation to Eq.~(\ref{scalingf}): $\mathcal{M}(V) = (n/v_T^d)\widetilde{\mathcal{M}}(c)$.

The decay rate for the field $A = \{ n,u_i,T\}$ reads $\xi_A^{(m)}$, where $m$ denotes the order in the Chapman-Enskog expansion. The corresponding dimensionless decay rate is
\begin{equation}
\xi_A^{(m)*} = \frac{\xi_A^{(m)}}{\nu_0}.
\end{equation}

%=========================================================
\section{Decay exponents of the homogeneous cooling state of PBA with hard sphere dynamics}\label{appendixdecayexponentshs}
%=========================================================
We shall recall results obtained in~\cite{coppexdroztrizac,coppex}. The density and temperature of the homogeneous cooling state for PBA of hard spheres are given by
\begin{subequations}
\begin{eqnarray}
n_H(t) &=& n_0 \left(1+p \frac{t}{t_0} \right)^{-\gamma_n^{\mathrm{S}}}, \label{stability1} \\
T_H(t) &=& T_0 \left(1+p \frac{t}{t_0} \right)^{-\gamma_T^{\mathrm{S}}}, \label{stability2}
\end{eqnarray}
\end{subequations}
where the decay exponents are $\gamma_n^{\mathrm{S}} = \xi_n^{(0)}(0) / t_0$, $\gamma_T^{\mathrm{S}} = \xi_T^{(0)}(0)/t_0$, and the relaxation time $t_0 = \xi_n^{(0)}(0) + \xi_T^{(0)}(0)/2$, where $\xi_n^{(0)}(0)$ and $\xi_T^{(0)}(0)$ are the decay rates at time $t=0$. The subscript $H$ denotes a quantity evaluated in the homogeneous state. Making use of the explicit values of $\xi_n^{(0)}(0)$ and $\xi_n^{(0)}(0)$ found in~\cite{coppexdroztrizac} and since $\gamma_v^{\mathrm{S}} = \gamma_T^{\mathrm{S}}/2$ one obtains
\begin{subequations}
\begin{eqnarray}
\gamma_n^{\mathrm{S}} &=& \frac{d+2}{4}\left(1-a_2 \frac{1}{16} \right) \nu_0 t_0,\\
\gamma_v^{\mathrm{S}} &=& \frac{1}{2} \frac{d+2}{8 d}\left(1+a_2 \frac{8d + 11}{16} \right) \nu_0 t_0,\\
t_0^{-1} &=& \left[ \frac{d+2}{4}\left(1-a_2 \frac{1}{16} \right) + \frac{d+2}{16 d}\left(1+a_2 \frac{8d + 11}{16} \right) \right] \nu_0,
\end{eqnarray}
\end{subequations}
where the kurtosis $a_2$ of the velocity distribution is~\cite{coppex}
\begin{equation}
a_2 = 8 \frac{3 - 2 \sqrt{2}}{4 d + 6 - \sqrt{2} + \frac{1-p}{p} 8 \sqrt{2} (d-1)}.
\end{equation}

%=========================================================
\section{Useful relations for the coefficients $\nu_\kappa^*$ and $\nu_\eta^*$}\label{appendix1}
%=========================================================
The expressions~(\ref{bigintegrals}) and~(\ref{lastone}) may be calculated with the help of the following relations. Let $X$ and $Y$ be arbitrary functions, $\mathcal{M}(\mathbf{V}) = n/(v_T^d \pi^{d/2}) \exp (- V^2/v_T^2)$ the Maxwellian in the scaling regime, then
\begin{equation}
\int_{\mathbb{R}^d} d \mathbf{v}_1 \, Y(\mathbf{v}_1) L_a[\mathcal{M} X] = \sigma^{d-1} \phi(x) v_T^{1-x} \int_{\mathbb{R}^{2d}} d \mathbf{v}_1 d \mathbf{v}_2 \,  v_{12}^x f^{(0)}(\mathbf{v}_1) \mathcal{M}(\mathbf{v}_2) X(\mathbf{v}_2)\left[ Y(\mathbf{v}_1) + Y(\mathbf{v}_2) \right], \label{lemme1}
\end{equation}
and
\begin{equation}
\int_{\mathbb{R}^d} d \mathbf{v}_1 \, Y(\mathbf{v}_1) L_c[\mathcal{M} X] = - \sigma^{d-1} \frac{\phi(x) v_T^{1-x}}{S_d} \int_{\mathbb{R}^{2d}} d \mathbf{v}_1 d \mathbf{v}_2 \, v_{12}^x f^{(0)}(\mathbf{v}_1) \mathcal{M}(\mathbf{v}_2) X(\mathbf{v}_2) \int d \widehat{\boldsymbol{\sigma}} (b-1) \left[ Y(\mathbf{v}_1) + Y(\mathbf{v}_2) \right], \label{lemme2}
\end{equation}
where $L_a g = - J_a[f^{(0)},g] - J_a[g,f^{(0)}]$ and $L_a g = - J_c[f^{(0)},g] - J_c[g,f^{(0)}]$ for an arbitrary function $g$. Let $\alpha \in \mathbb{R}^+$, then
\begin{eqnarray}
\int_{\mathbb{R}^d} d \mathbf{x} |\mathbf{x}|^n \mathrm{e}^{-\alpha x^2} &=& \frac{\pi^{d/2}}{\alpha^{(d+n)/2}} \frac{\Gamma[(d+n)/2]}{\Gamma(d/2)}, \label{app3eq5}\\
\int_{\mathbbm{R}^d} d \mathbf{x} \, |\mathbf{x}|^n \, \mathrm{e}^{-\alpha \mathbf{x}^2} x_i x_j &=& \frac{\pi^{d/2}}{\alpha^{(d+n+2)/2}} \frac{d+n}{2d} \frac{\Gamma\left[(d+n)/2\right]}{\Gamma\left[d/2\right]} \delta_{ij}.
\end{eqnarray}
In the integrals below, the results when $\theta(\widehat{\boldsymbol{\sigma}} \cdot \mathbf{g})$ is absent are obtained upon multiplying the value of $\beta_n$ by two. Finally for $\widehat{\boldsymbol{\sigma}} = (\sigma_1, \ldots, \sigma_d)$, $\mathbf{g} \in \mathbbm{R}^d$, $|\widehat{\boldsymbol{\sigma}}| = 1$, $|\widehat{\mathbf{g}}|=1$, we have
\begin{eqnarray}
& & \int d \widehat{\boldsymbol{\sigma}} \, \theta(\widehat{\boldsymbol{\sigma}} \cdot \mathbf{g}) (\widehat{\boldsymbol{\sigma}} \cdot \mathbf{g})^n \sigma_i \sigma_j = \frac{\beta_n}{n+d} g^{n-2} (n g_i g_j + g^2 \delta_{ij}),\label{finally1}\\
& & \int d \widehat{\boldsymbol{\sigma}} \, \theta(\widehat{\boldsymbol{\sigma}} \cdot \mathbf{g}) (\widehat{\boldsymbol{\sigma}} \cdot \mathbf{g})^n \sigma_i = \beta_{n+1} g^{n-1} g_i,\label{finally2}\\
& & \beta_n = \int d \widehat{\boldsymbol{\sigma}} \, \theta(\widehat{\boldsymbol{\sigma}} \cdot \widehat{\mathbf{g}}) (\widehat{\boldsymbol{\sigma}} \cdot \widehat{\mathbf{g}})^n = \pi^{(d-1)/2} \frac{\Gamma\left[ (n+1)/2 \right]}{\Gamma\left[ (n+d)/2 \right]}.\label{finally3}
\end{eqnarray}

%=========================================================
\section{Exact relations for the transport coefficients of the Maxwell model}\label{appendix2}
%=========================================================

Following the same route as in~\cite{coppexdroztrizac} we may rewrite the right hand side of Eq.~(\ref{boltzmann1}) such that
\begin{equation}
[ \partial_t^{(0)} + J ] f^{(1)} = A_i \nabla_i \ln T + B_i \nabla_i \ln n + C_{ij} \nabla_i u_j + p \Omega f^{(1)},\label{boltzexacta}
\end{equation}
where
\begin{subequations}
\begin{eqnarray}
A_i &=& \frac{V_i}{2} \frac{\partial}{\partial V_j} [V_j f^{(0)}] - \frac{k_B T}{m} \frac{\partial f^{(0)}}{\partial V_i}, \label{coeff1} \\
B_i &=& - V_i f^{(0)} - \frac{k_B T}{m} \frac{\partial f^{(0)}}{\partial V_i}, \label{coeff2} \\
C_{ij} &=& \frac{\partial}{\partial V_i}[V_j f^{(0)}] - \frac{1}{d} \frac{\partial}{\partial V_k}[ V_k f^{(0)} ]\delta_{ij}, \label{coeff3}
\end{eqnarray}\label{coeff}
\end{subequations}
and $\Omega$ is a linear operator defined by
\begin{equation}
\Omega g = f^{(0)} \xi_n^{(1)}[f^{(0)},g] - \frac{\partial f^{(0)}}{\partial V_i} v_T \xi_{u_i}^{(1)}[f^{0},g] + \frac{\partial f^{(0)}}{\partial T} T \xi_T^{(1)}[f^{(0)},g].
\label{defOmega}
\end{equation}
The quantity $g$ is either $\mathcal{A}_i$, $\mathcal{B}_i$, or $\mathcal{C}_{ij}$, and the functionals $\xi_n^{(1)}$, $\xi_{u_i}^{(1)}$, and $\xi_T^{(1)}$ are obtained from Eqs.~(\ref{decay1}) upon replacing $f^{(1)}$ by $g$.

\subsection{Pressure tensor}
Integrating the Boltzmann equation~(\ref{boltzexacta}) over $\mathbf{V}$ with weight $m V_i V_j$ and taking into account the symmetry properties of the coefficients~(\ref{coeff}) one obtains
\begin{equation}
\partial_t^{(0)} P_{ij}^{(1)}(\mathbf{r},t) + p \int_{\mathbbm{R}^d} d \mathbf{V} \, m V_i V_j L_a[f^{(0)},f^{(1)}] + (1-p) \int_{\mathbbm{R}^d} d \mathbf{V} \, m V_i V_j L_c[f^{(0)},f^{(1)}] = \int_{\mathbbm{R}^d} d \mathbf{V} \, m V_i V_j C_{kl}(\mathbf{V}) \nabla_k u_l, \label{eqexact}
\end{equation}
where we have made use of the definition~(\ref{pressure1}) for the pressure tensor. The same definition further allows us to write
\begin{equation}
\int_{\mathbbm{R}^d} d \mathbf{V} \, m V_i V_j L_a[f^{(0)},f^{(1)}] = \xi_n^{(0)} P_{ij}^{(1)}(\mathbf{r},t), \label{eqexact1}
\end{equation}
and using additionally Eq.~(\ref{lemme2}), and~(\ref{finally1}) to~(\ref{finally3}):
\begin{equation}
\int_{\mathbbm{R}^d} d \mathbf{V} \, m V_i V_j L_c[f^{(0)},f^{(1)}] = \xi_n^{(0)} \frac{2}{d+2} P_{ij}^{(1)}(\mathbf{r},t). \label{eqexact2}
\end{equation}
Finally
\begin{equation}
\int_{\mathbbm{R}^d} d \mathbf{V} \, m V_i V_j C_{kl}(\mathbf{V}) \nabla_k u_l = - p^{(0)} \Delta_{ijkl} \nabla_k u_l, \label{eqexact3}
\end{equation}
where
\begin{equation}
\Delta_{ijkl} = \delta_{ik} \delta_{jl} + \delta_{jk} \delta_{il} - \frac{2}{d} \delta_{ij} \delta_{kl}.
\end{equation}
Insertion of Eqs.~(\ref{eqexact1}) to~(\ref{eqexact3}) 
in~(\ref{eqexact}) yields
\begin{equation}
\left[\partial_t^{(0)} + p \xi_n^{(0)} + (1-p)\xi_n^{(0)} \frac{d+2}{2} \right] P_{ij}^{(1)}(\mathbf{r},t) = - p^{(0)} \Delta_{ijkl} \nabla_k u_l. \label{eqexact-2}
\end{equation}
The solution of Eq.~(\ref{eqexact-2}) is $P_{ij}^{(1)} = - \eta \Delta_{ijkl} \nabla_k u_l$. Functional dependence analysis shows that $\eta \propto T^{1/2}$, and since to zeroth order the temperature is conserved $\partial_t P_{ij}^{(1)} = 0$. Eq.~(\ref{eqexact-2}) thus gives
\begin{equation}
\eta^* = \frac{1}{p \frac{d+2}{2} + (1-p)}.
\end{equation}

\subsection{Heat flux}
Integrating the Boltzmann equation~(\ref{boltzexacta}) over $\mathbf{V}$ with weight $m V^2 V_i / 2$ and taking into account the symmetry properties of the coefficients~(\ref{coeff}) one obtains
\begin{multline}
\partial_t^{(0)} q_i^{(1)}(\mathbf{r},t) + p \int_{\mathbbm{R}^d} d \mathbf{V} \, \frac{1}{2} m V^2 V_i L_a[f^{(0)},f^{(1)}] + (1-p) \int_{\mathbbm{R}^d} d \mathbf{V} \, \frac{1}{2} m V^2 V_i L_c[f^{(0)},f^{(1)}] \\
= \int_{\mathbbm{R}^d} d \mathbf{V} \, \frac{1}{2} m V^2 V_i  A_{k}(\mathbf{V}) \nabla_k \ln T + \int_{\mathbbm{R}^d} d \mathbf{V} \, \frac{1}{2} m V^2 V_i  B_{k}(\mathbf{V}) \nabla_k \ln T, \label{eqexact--2}
\end{multline}
where we have made use of the definition~(\ref{heat}) for the heat flux to first order. Moreover, one finds
\begin{equation}
\int_{\mathbbm{R}^d} d \mathbf{V} \,  \frac{1}{2} m V^2 V_i  L_a[f^{(0)},f^{(1)}] = \xi_n^{(0)} q_{i}^{(1)}(\mathbf{r},t), \label{eqexact1--2}
\end{equation}
and using additionally Eq.~(\ref{lemme2}), and~(\ref{finally1}) to~(\ref{finally3}):
\begin{equation}
\int_{\mathbbm{R}^d} d \mathbf{V} \, \frac{1}{2} m V^2 V_i L_c[f^{(0)},f^{(1)}] = \frac{2(d-1)}{d(d+2)} \xi_n^{(0)} q_{i}^{(1)}(\mathbf{r},t). \label{eqexact2--2}
\end{equation}
Finally
\begin{equation}
\int_{\mathbbm{R}^d} d \mathbf{V} \, \frac{1}{2} m V^2 V_i A_{k}(\mathbf{V}) \nabla_k \ln T = - \frac{d+2}{2} \frac{p^{(0)} k_B}{m} \nabla_i T, \label{eqexact3--2}
\end{equation}
\begin{equation}
\int_{\mathbbm{R}^d} d \mathbf{V} \, \frac{1}{2} m V^2 V_i B_{k}(\mathbf{V}) \nabla_k \ln n = 0. \label{eqexact3b--2}
\end{equation}
Insertion of Eqs.~(\ref{eqexact1--2}) to~(\ref{eqexact3b--2}) 
in~(\ref{eqexact--2}) gives
\begin{equation}
\left[\partial_t^{(0)} + p \xi_n^{(0)} + (1-p) \xi_n^{(0)} \frac{2(d-1)}{d(d+2)} \right] q_{i}^{(1)}(\mathbf{r},t) = - \frac{d+2}{2} \frac{p^{(0)} k_B}{m} \nabla_i T. \label{eqexact-2--2}
\end{equation}
The solution of Eq.~(\ref{eqexact-2--2}) is $q_{i}^{(1)} = - \lambda \nabla_i T - \mu \nabla_i n$. Functional dependence analysis shows that $\lambda \propto T^{1/2}$ and $\mu \propto T^{3/2} n^{-1}$, therefore $\partial_t q_{i}^{(1)} = p \xi_n^{(0)} \mu \nabla_i n$. In order to satisfy Eq.~(\ref{eqexact-2--2}) it is therefore required that $\mu = 0$ and
\begin{equation}
\lambda^* = \frac{1}{p \frac{d(d+2)}{2(d-1)} + (1-p)}.
\end{equation}

%=========================================================

\end{document}